\documentclass{amsart}
\usepackage{amssymb}
\usepackage[T1]{fontenc}
\usepackage[british]{babel}
\usepackage{graphicx}
\usepackage[utf8]{inputenc}
\usepackage{mathrsfs}
\usepackage{xspace}

\newcommand{\RR}{\ensuremath{\mathbb R}\xspace}
\renewcommand{\S}[1]{\ensuremath{\mathbb{S}^#1}\xspace}
\newcommand{\scri}{{\mathscr I}}
\newcommand{\dd}{\mathbf{d}}
\newcommand{\del}{\partial}

\newcommand{\eps}{\varepsilon}
\newcommand{\thorn}{\mbox{\th}}
\newcommand{\thorp}{\mbox{\th}'}
\newcommand{\etp} {\eth'}
\newcommand{\Y}[2]{{}_{#1}Y_{#2}}
\DeclareMathOperator{\ArcTanh}{atanh}

\graphicspath{{../ArticleSubmission/}}

\title[Numerical space-times near space-like and null infinity]{Numerical space-times near space-like and null infinity. The spin-2 system on Minkowski space}

\author[F. Beyer]{Florian Beyer}
\address{Department of Mathematics and Statistics, University of Otago, PO Box 56, Dunedin 9010, New Zealand}
\email{fbeyer@maths.otago.ac.nz}

\author[G. Doulis]{Georgios Doulis}
\address{Department of Mathematics and Statistics, University of Otago, PO Box 56, Dunedin 9010, New Zealand}
\email{gdoulis@maths.otago.ac.nz}

\author[J. Frauendiener]{J\"org Frauendiener}
\address{Department of Mathematics and Statistics, University of Otago, PO Box 56, Dunedin 9010, New Zealand}
\email{joergf@maths.otago.ac.nz}

\thanks{This research was funded by the Marsden Fund of the Royal Society of New Zealand}

\author[B. Whale]{Ben Whale}
\address{Department of Mathematics and Statistics, University of Otago, PO Box 56, Dunedin 9010, New Zealand}
\email{bwhale@maths.otago.ac.nz}
\date{\today}

\begin{document}

\begin{abstract}
  In this paper we demonstrate for the first time that it is possible to solve numerically the Cauchy problem for the linearisation of the general conformal field equations near  spacelike infinity, which is only well-defined in Friedrich's cylinder picture. We have restricted ourselves here to the ``core'' of the equations -- the spin-2 system -- propagating on Minkowski space. We compute the numerical solutions for various classes of initial data, do convergence tests and also compare to exact solutions. We also choose initial data which intentionally violate the smoothness conditions and then check the analytical predictions about singularities. This paper is the first step in a long-term investigation of the use of conformal methods in numerical relativity.
\end{abstract}
\maketitle

\section{Introduction}
\label{sec:introduction}

The subject of numerical relativity has made significant strides since 2005, when Pretorius~\cite{Preto2005:Evolution-of-Binary-Black-Hole} presented the first simulation of a binary black hole system, which extended over several orbits. Since then the numerical relativity community has developed more and more sophisticated codes to track the evolution of highly relativistic binaries of compact objects, increasingly including neutron stars. The codes are stable and accurate so that they can be used for production runs. In fact, the problem nowadays is not so much within an individual run but lies in the vastness of the parameter space (eight dimensional in the case of two spinning black holes). This calls for the development of sophisticated methods to exhaust this space~\cite{FieldGalle2011:Reduced-Basis-Catalogs}.

However, there are still several problems with the numerical simulations, which have not been entirely solved. We mention here only the issue of the outer boundary and the feasibility of a global numerical simulation of an entire space-time. These are largely conceptual issues, which have been dealt with mostly by practical considerations. The background behind these issues is that all numerical simulations of binary systems are based on an \emph{idealisation}: one assumes that the system under consideration does not interact with anything else, i.e., that it be an \emph{isolated system}~\cite{geroch77:_asym_struc,frauendiener2004:_confor}. 

Mathematically, this idealisation leads to the requirement that the space-time manifold describing the system be asymptotically flat. The main consequence of this is that the gravitational radiation, which is emitted by the system can be read off uniquely at null-infinity, commonly denoted by $\scri$. This is a non-trivial result since the background independence of the theory (a consequence of the general coordinate invariance) prohibits the introduction of any absolute structure to which the gravitational waves (the proverbial `ripples' on space-time) can be referred to. Furthermore, the `infinity' of an asymptotically flat space-time is unique. As a consequence gravitational radiation is a gauge dependent concept except at infinity. Note, that infinity in the space-time context refers to infinite large distances as well as infinitely long times. So, the unique gravitational wave information from a radiating system can be detected only at infinite distances from the system after infinitely long times.

Obviously, this imposes a huge burden on the numerical simulations, which cannot incorporate infinite resources and, therefore, need to recourse to practical measures. This is done by imposing an outer boundary, which is considered to be `practically' at infinity and the gravitational field is traced along the time-like hyper-surface swept out by the boundary during the simulation. The information contained in the field near the boundary is an important ingredient in very elaborate  `extraction schemes' which are designed to provide an approximation to the radiation field at infinity.

The introduction of an artificial boundary into an infinite problem usually comes with several problems. Not only is it necessary to specify boundary conditions but these should be physically reasonable and, in addition, should lead to a mathematically well-posed initial boundary value problem (IBVP) and a numerically stable evolution scheme. The issue of the introduction of a time-like artificial boundary is physically relevant, as was pointed out by Zenginoglu~\cite{Zenginoglu:2008ch}, since the decay rates of the tails are different on time-like boundaries compared to null-infinity. So one cannot simply replace one with the other. In any case, introducing an artificial boundary is \emph{always} an additional approximation beyond the conceptual idealisation of an isolated system. 

Experience shows that for all practical purposes the current codes are good enough. However, from a puristic point of view the situation is still unsatisfactory. In this paper we will present the first step of a numerical approach, which will ultimately allow us to evolve space-times globally from asymptotically Euclidean initial data. Our method is based on the geometric description of asymptotically simple space-times developed by Penrose and their analytical treatment pioneered by Friedrich.

In 1965, Penrose~\cite{penrose64:_light_cone_inf,Penrose:1965p3324} had realised that asymptotically flat (and, more generally, asymptotically simple) space-times can be characterised by the fact that they have a conformal boundary, i.e., they can be isometrically embedded into a larger Lorentzian manifold after a conformal rescaling of their metric such that the boundary of the embedded manifold is given by the zero-set of the conformal factor. This means that these boundary points must be regarded as being at infinite distance from any event in the original `physical' space-time.

Of course, such a geometric characterisation does not imply anything about the existence of general solutions of the field equations, which do in fact admit such a conformal boundary. This gap has been filled by Friedrich~\cite{Friedrich:1983vx,Friedrich:1986eo} using what he called the conformal field equations (CFE), a slight generalisation of Einstein's field equations. One of his main results was the theorem on the well-posedness of the hyperboloidal initial value problem. This means that if appropriate data on a hyperboloidal initial hyper-surface are given then there exists a solution of the CFE at least for some time in the future of the initial hyper-surface. This result implies that the corresponding physical space-time admits a smooth conformal boundary, i.e., it is asymptotically flat. If the initial data are close to Minkowski data then the solution exists for such a long time that the physical space-time is in fact future complete, i.e., it exists for all times. 

This semi-global existence result has been combined with a result by Corvino~\cite{corvino00:_scalar_einst} on the existence of asymptotically Euclidean initial data with an exact Schwarzschild exterior by Chru\'sciel and Delay~\cite{chruscieldelay02:_exist} to obtain a global existence theorem for asymptotically flat space-times. Thus, Penrose's conjecture on the existence of general asymptotically flat space-times has finally been settled to the affirmative.

The CFE operate on the larger, conformally related `unphysical' manifold, within which the original `physical' space-time occupies a finite set. This property makes it very tempting to use the CFE for numerical simulations because they capture the entire physical space-time without the need for an artificial boundary. The feasibility of this idea has been shown in several different scenarios~\cite{stewart89:_numer_iii,huebner94:_method_calc_sing_spti,huebner96:_numer_glob,Frauendiener:1998vo,Frauendiener:1998th,Frauendiener:1999vl,huebner01:_from_now,Beyer:2008gg,Beyer:2009vw,Beyer:2009ta,Frauendiener:2002ux,Zenginoglu:2007wo}. The main advantage of this approach is, of course, the fact that the global physical space-time is covered by one computational domain. It is worth mentioning that we are dealing here with a conformal rescaling of the metric, so that all the causal relations and the propagation properties of the (gravitational) waves are identical in the physical and the conformal manifolds, see~\cite{Frauendiener:2000vf}. Furthermore, since the conformal boundary is accessible from the conformal space-time this implies that the unique radiation data produced by the system can be read off directly on the boundary without any additional approximation.

The main drawback of this approach (apart from the increased number of unknowns) has been the fact that the conformal factor itself is a variable in the system and, therefore, evolves. Hence, its zero-set is not known beforehand (unless one employs a particular shift gauge called `$\scri$-freezing'~\cite{Frauendiener:1998th}). It has to be located within the computational domain at each time-step in order to determine the radiation data.

Another issue has been the fact that initial data need to be specified on hyperboloidal hyper-surfaces. This has the consequence that only \emph{pre}diction is possible but not \emph{retro}diction. Also within this hyperboloidal initial value problem it is not possible to study effects of gravitational radiation coming in from null-infinity, such as for example the scattering of gravitational waves. For such questions it is necessary to include the region around space-like infinity, the avoidance of which has been the motivation for introducing hyperboloidal hyper-surfaces in the first place.

We should point out that there exist other numerical approaches which are based on hyperboloidal hyper-surfaces but try to avoid the use of the full conformal field equations. In~\cite{Zenginoglu:2007wo} Zenginoglu describes a setup which regards the conformal factor as an external matter field coupled to the  Einstein equations in the conformal manifold. The energy-momentum tensor constructed from the conformal transformation properties of the Einstein tensor is singular on $\scri$ but there are ways to make it regular. Zenginoglu shows that in spherical symmetry one can evolve a reduced system using the so called generalised harmonic gauge. He also points out that in this approach the conformal factor is a free function and can be chosen so that it is identically equal to one inside a certain compact region, which makes the inclusion of real matter fields easier.

Another attempt at including null-infinity without using the conformal field equations is based on the work by Moncrief and Rinne~\cite{Moncrief:2009ds}. They employ a Hamiltonian framework based on conformally compactified hyperboloidal hyper-surfaces and show that the Hamiltonian and momentum constraints, which are formally singular on $\scri$, are in fact regular and can be evaluated there. They use a maximally constrained formulation for the evolution. First numerical results have been presented in~\cite{Rinne:2010ew}.

In~\cite{Friedrich:1998tc} Friedrich describes another set of conformal field equations, the generalised CFE (in contrast to the earlier `metric' CFE), which can be used to study exactly the region near space-like infinity. These GCFE make full use of the conformal geometry not only on the level of the metric but also on the level of the connection. This has the important consequence that it is now possible to fix the conformal factor beforehand as a function of the coordinates so that the coordinate location of $\scri$ is known at every instant. Friedrich also shows that one can `blow up' space-like infinity so that it is represented by a cylinder connecting past and future null-infinity.

The main advantage of this new system of conformal field equations is the fact that one can employ a gauge, such that almost all the equations take the form of simple ODEs along the time-like coordinate vector. The only equations not of this type provide the evolution of the rescaled Weyl curvature, what Penrose called the gravitational field. This system is obtained from the Bianchi identities and it is easily shown to be symmetric hyperbolic. The cylinder is a total characteristic for the GCFE, i.e., the equations reduce to an intrinsic system of evolution equations on the cylinder, emphasising the sole purpose of the cylinder, namely to provide a connection between the very late ingoing radiation and the very early outgoing radiation.

These properties of the GCFE call out for an attempt to use them for numerical simulations. The fixed conformal factor, the ensuing fixed location of $\scri$ in the computational domain, the fact that one can cover a neighbourhood of space-like infinity including and beyond $\scri^-$ and $\scri^+$ and the advantageous form of the evolution system are all too tempting. In fact, the first numerical application of the GCFE has been carried out by Zenginoglu~\cite{Zenginoglu:2007wo} (an even earlier application of the GCFE to a different setting has been considered in \cite{Beyer:2007vp}). He was able to compute a massless axisymmetric radiative space-time in the neighbourhood of spatial infinity. 

However, all this comes at a price. The system of equations intrinsic to the cylinder degenerates on the locations, where the cylinder meets $\scri$. This poses serious analytical problems because generic solutions will develop singularities there, which are expected to travel along null-infinity, thereby spoiling its smoothness and hence the concept of gravitational radiation. In addition, the numerical simulations might be affected by the degeneracies in the evolution equations.

In order to get a feeling for the possible effects we start out in this paper with a simplified situation. We study the gravitational perturbations on Minkowski space-time near space-like infinity in the gauge, which exhibits the cylinder described above. This amounts to solving the linear spin-2 zero-rest-mass field equation on a particular conformally flat space-time. However, as pointed out above, this system already captures the essential properties of the full non-linear system. The equations show the exact same behaviour as the fully non-linear GCFE in the sense that they reduce to an intrinsic set of evolution equations on the cylinder and they degenerate on the intersection with null-infinity. Our goal is to find out how the numerics reacts to the above mentioned degeneracies and how one can control the solutions at these locations.

The paper is organised as follows. In sect.~\ref{sec:spin-2-system} we present the system under study. In sect.~\ref{sec:analyt-prop} we describe some analytical properties of the system which are relevant for the solution. Sect.~\ref{sec:numer-impl} contains the description of the numerical implementation and presents our results. We close the paper with a discussion in sect.~\ref{sec:discussion}.

\section{The spin-2 system near space-like infinity}
\label{sec:spin-2-system}

In this section we derive the spin-2 zero-rest-mass equation on Minkowski space-time near space-like infinity. This section follows largely the exposition in the work by Friedrich~\cite{Friedrich:1998tc,Friedrich:2003hp}.

\subsection{Minkowski space-time near spatial infinity}
\label{sec:minkowski-space-time}

The metric of Minkowski space-time $\tilde M$ in Cartesian coordinates $X^a$ is given by
\begin{equation}
  \label{eq:1}
  \tilde g = \eta_{ab} \dd X^a \dd X^b = \dd T^2 - \dd R^2 - R^2 \dd \omega^2
\end{equation}
where, $\eta_{ab}=\mathrm{diag}(1,-1,-1,-1)$, $T=X^0$, $R^2 = (X^1)^2 + (X^2)^2 + (X^3)^2$ and $\dd\omega^2$ is the unit sphere metric. The exterior $\tilde M_-$ of the null-cone of the origin is given by $X^aX_a<0$. In order to focus on space-like infinity we now perform a reflection at the origin on $\tilde M_-$ by defining
\[
X^a = - \frac{x^a}{x\cdot x}, \qquad x\cdot x :=\eta_{ab}x^a x^b = \frac1{(X\cdot X)}.
\]
This puts the metric into the form
\begin{equation}
  \label{eq:2}
  \tilde g = \frac{\eta_{ab}\dd x^a \dd x^b}{(x\cdot x)^2}.
\end{equation}
This metric is singular whenever $x\cdot x=0$, i.e., on the null-cone at infinity. So we define a conformally related metric
\begin{equation}
  \label{eq:3}
  g' = \Omega^2 \tilde g = \eta_{ab}\dd x^a \dd x^b , \qquad \Omega = -(x\cdot x)
\end{equation}
which extends smoothly to the null-cone of infinity.
Note, that space-like infinity with respect to $\tilde M_-$ is represented in this metric as the point $x^a=0$. In order to exhibit the cylindrical structure referred to above Friedrich~\cite{Friedrich:1998tc} performs a further rescaling of the metric using a function $\kappa(r) = r \mu(r)$ where $r^2 =  (x^1)^2 + (x^2)^2 + (x^3)^2$ and $\mu$ is a smooth function with $\mu(0)=1$. The rescaled metric is defined by
\begin{equation}
  \label{eq:4}
  g = \frac1{\kappa^2} g' = \frac{\Omega^2}{\kappa^2} \tilde g = \frac1{\kappa^2} \left( \dd (x^0)^2 - \dd r^2 - r^2 \dd \omega^2\right).
\end{equation}
The final step is the introduction of a new time-coordinate $t$ by $x^0 = \kappa(r) t$. This gives the final form of the metric
\begin{equation}
  \label{eq:5}
  g = \frac1{\kappa^2} \left(\kappa^2 \dd t^2 + 2 t \kappa \kappa' \dd t \dd r - (1-t^2 \kappa'{}^2) \dd r^2 - r^2 \dd \omega^2 \right).
\end{equation}
Note, that this metric is spherically symmetric.

\subsection{The cylinder at space-like infinity}
\label{sec:cylinder-at-space}

We write the metric~\eqref{eq:5} in the form
\begin{equation}
  \label{eq:10}
  g =  \dd t^2 + 2 t \frac{r\kappa'}{\kappa} \dd t \frac{\dd r}r - \frac{(1-t^2 \kappa'{}^2)}{\mu^2} \left(\frac{\dd r}{r}\right)^2 - \frac1{\mu^2} \dd \omega^2.
\end{equation}
This shows that the surfaces $(t,r)=\mathrm{const}$ have a non-vanishing area for each value of $t$ and all $r>0$. Since $\mu(0)=1$ this also holds for $r=0$ by continuity. 
In the coordinates $(t,r)$ the set $r=0$ is a cylinder with topology $\RR\times\S{2}$.

The conformal factor used between the Minkowski metric~\eqref{eq:1} and the metric~\eqref{eq:5} is
\[
\Theta= \frac{\Omega}{\kappa} = \frac{r^2 - t^2\kappa^2}{r\mu(r)} = \frac{r}{\mu} (1 - t^2\mu^2).
\]
\begin{figure}[htb]
  \centering
  \includegraphics[height=8cm]{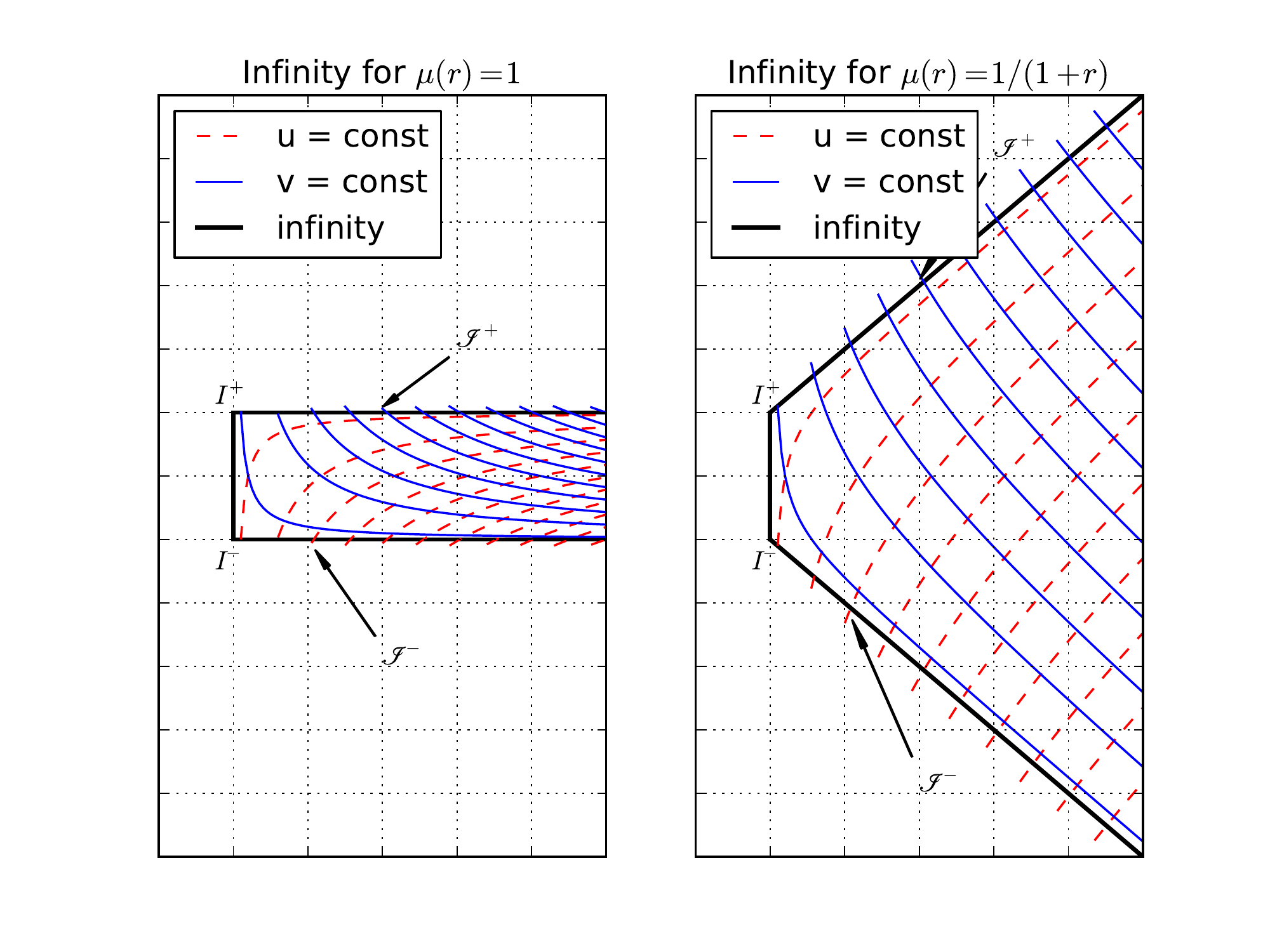}
  \caption{The neighbourhood of space-like infinity in two different $(t,r)$-coordinate representations corresponding to $\mu(r)=1$ and $\mu(r)=1/(1+r)$}
  \label{fig:cylinder}
\end{figure}

Let $M_-=\{\Theta>0\}$ denote the part of the conformal space-time, which corresponds to Minkowski space-time~$\tilde M$. The boundary of $M_-$ consists of three parts: we have $\scri^\pm = \{1 \mp t\mu = 0\}$ and we denote by $I$ the set $\{ -1 < t < 1, r = 0\}$. Furthermore, let $I^\pm = \{ t= \pm1, r=0\}$ denote the regions, where $I$ touches $\scri^\pm$, and $I^0=\{t=r=0\}$ the intersection of the hyper-surface $t=0$ with $I$. We choose the function $\mu$ to be of the form
\[
\mu(r) = \frac1{1 + n r}
\]
and we will use mainly two choices for $n$, namely $n=0$ ($\mu=1$) or $n=1$ ($\mu=1/(1+r)$). For the first choice, $\scri$ is located at $t=\pm1$, while the second choice puts it onto $t=\pm(1+r)$, see Fig.~\ref{fig:cylinder}.

In order to get some insight into the structure near space-like infinity we introduce double null coordinates 
\begin{equation}
\label{eq:8}
u = \kappa t - r, \qquad v = \kappa t + r,
\end{equation}
which puts the metric into the form
\[
g = \frac1{\kappa^2}\dd u \dd v - \frac1{\mu^2}\dd\omega^2.
\]
Null-infinity is characterised by the vanishing of one of the null coordinates, the other one being non-zero ($u=0$, $v\ne0$ on $\scri^+$ and $u\ne0$, $v=0$ on $\scri^-$). Both coordinates vanish on the cylinder~$I$. In Fig.~\ref{fig:cylinder} we also display the lines of constant $u$ and $v$. Since these are coordinates adapted to the conformal structure of $M_-$ we can see clearly, that the cylinder is `invisible' from the point of view of the conformal structure. This structure is exactly the same as the representation of space-like infinity in the coordinates $x^a$ on $M_-$, it appears like a point. However, the differentiable structures defined by the $(u,v)$ and the $(t,r)$ coordinates are completely different near the boundary~$r=0$.

\subsection{The spin-2 equation}
\label{sec:spin-2-equation}
We use the spinorial form for the spin-2 zero-rest-mass field $\phi_{ABCD} = \phi_{(ABCD)}$, see Penrose and Rindler~\cite{Penrose:1984wr}, the conventions of which  (the GHP-formalism) we adopt throughout this work. The field equation is
\begin{equation}
  \label{eq:6}
  \nabla^A{}_{A'} \phi_{ABCD} = 0,
\end{equation}
where $\nabla_{AA'}$ is the spinorial connection defined by a general $3+1$ metric $g$. Using the compacted spin-coefficient formalism~\cite{Geroch:1973wg,Penrose:1984wr} and defining the spinor field components $\phi_k$ in the usual way ($\phi_0=\phi_{0000}$, $\phi_1=\phi_{0001}$,\ldots) we obtain the equations\footnote{In the following equation only, the symbol $\kappa$ stands for the spin coefficient from the GHP formalism and is not to be confused with the function $\kappa$ introduced before (for the metric $g$ in Eq.~\eqref{eq:10}). Since, as will see shortly, the spin coefficient $\kappa$ vanishes identically under our assumptions, we continue to use the symbol $\kappa$ for the function  in Eq.~\eqref{eq:10}. Moreover, it should always be clear from the context whether a prime corresponds to the operator from the GHP formalism or to a radial derivative as before.} for $k=1:4$
\begin{equation}
  \label{eq:7}
\begin{aligned}
\thorn \phi_k - \etp \phi_{k-1} &= - k \tau' \phi_{k-1} - (4 - k) \kappa \phi_{k+1}  + (k-1) \sigma' \phi_{k-2}  + (5-k) \rho \phi_k, \\ 
\thorp \phi_{k-1} - \eth \phi_{k} &= - (5-k) \tau \phi_{k} - (k-1) \kappa' \phi_{k-2}  + (4-k) \sigma \phi_{k+1} + k \rho' \phi_{k-1}. 
\end{aligned}
\end{equation}

The metric $g$ in Eq.~\eqref{eq:10} is spherically symmetric. Using a null-tetrad, which is adapted to this symmetry, we know immediately that all spin-coefficients with non-zero spin-weight vanish. The only non-vanishing spin-coefficients in Eq.~\eqref{eq:7} are 
\[
\rho=-\rho' = \frac1{\sqrt2} r \mu'.
\] 
We choose the real null-vectors of the null-tetrad as
\begin{equation}
 \label{null_tetrad}
 l^a\del_a = \frac1{\sqrt2}\left(\left(1 - t \kappa'\right) \del_t + \kappa \del_r  \right), \qquad
 n^a\del_a = \frac1{\sqrt2}\left(\left(1 + t \kappa'\right) \del_t - \kappa \del_r  \right),
\end{equation}
and complement them with a complex space-like null-vector $m^a\del_a$, which is tangent to the spheres of symmetry; one particular choice is
\[m^a \del_a= \frac{\mu}{\sqrt2} \left( \del_\theta - \frac{i}{\sin\theta}\, \del_\phi\right),\]
where $(\theta,\phi)$ are the standard polar coordinates on $\S{2}$.
Then the relationship between the GHP-operators $\thorn$ and $\thorp$ and the directional derivatives $D$ and $D'$ (see~\cite{Penrose:1984wr}) are mediated by the unweighted spin-coefficients $\eps$ and $\gamma$, which are 
\[
\eps=\gamma= -\frac1{2\sqrt2} \kappa',
\]
for the metric $g$. Furthermore, we redefine the $\eth$ and $\etp$ operators so that they refer to the unit sphere. This is achieved by the replacement
\begin{equation}
\eth \mapsto \frac\mu{\sqrt2} \eth, \qquad
\etp \mapsto \frac\mu{\sqrt2} \etp;\label{eq:ethreplacement}
\end{equation}
these new operators act on a function $\eta$, which is defined on the unit sphere and which has spin-weight $s$, by
\[\eth\eta=m^a \del_a\eta-s\cot\theta\, \eta, \quad
\etp\eta=\bar m^a \del_a\eta-s\cot\theta\, \eta.\]
With these specialisations the eqs.~\eqref{eq:7} become
\begin{equation}
  \label{eq:9}
\begin{aligned}
(1-t\kappa') \del_t \phi_k + \kappa \del_r \phi_k - (3\kappa'  - (5-k) \mu) \phi_k &=  \mu\etp \phi_{k-1} , &k=1:4,\\
(1+t\kappa') \del_t \phi_{k} - \kappa \del_r \phi_{k} + (3\kappa' + (k+1) \mu ) \phi_{k}  &= \mu \eth \phi_{k+1}, &k=0:3.
\end{aligned}
\end{equation}
Finally, we use the spherical symmetry of the background metric to expand the spinor component $\phi_k$ into spin-weighted spherical harmonics $\Y{s}{lm}$ with $s=2-k$, $|s| \le l$ and $|m| \le l$, i.e., we write for any $(t,r,\mathbf{e}) \in \mathbb{R}\times\mathbb{R}^+\times S^2$
\[
\phi_k(t,r,\mathbf{e}) = \sum_{l=|s|}^\infty \sum_{m=-l}^l \phi_k^{lm}(t,r)\, \Y{s}{lm}(\mathbf{e}).
\]
Using the properties of the unit-sphere $\eth$ and $\etp$ operators acting on spin-weighted spherical harmonics
\[
\eth \Y{s}{lm} = -\sqrt{l(l+1) - s (s+1)}\; \Y{s+1}{lm},\qquad 
\etp \Y{s}{lm} = \sqrt{l(l+1) - s (s-1)}\; \Y{s-1}{lm},
\]
we can decouple the equations and obtain a system of PDEs in $1+1$ dimensions for any admissible pair $(l,m)$ 
\begin{equation}
\begin{aligned}
\label{eq:fullminus}
(1-t\kappa') \del_t \phi_1^{lm} + \kappa \del_r \phi_1^{lm} &=  (3\kappa' - 4 \mu) \phi_1^{lm} +  \mu\alpha_2 \phi_0^{lm} ,\\
(1-t\kappa') \del_t \phi_2^{lm} + \kappa \del_r \phi_2^{lm} &=  (3\kappa' - 3 \mu) \phi_2^{lm} +  \mu \alpha_0 \phi_1^{lm} ,\\
(1-t\kappa') \del_t \phi_3^{lm} + \kappa \del_r \phi_3^{lm} &=  (3\kappa' - 2 \mu) \phi_3^{lm} +  \mu \alpha_0 \phi_2^{lm} ,\\
(1-t\kappa') \del_t \phi_4^{lm} + \kappa \del_r \phi_4^{lm} &=  (3\kappa' -  \mu) \phi_4^{lm} +  \mu \alpha_2 \phi_3^{lm} ,
\end{aligned}
\end{equation}
and
\begin{equation}
\label{eq:fullplus}
\begin{aligned}
(1+t\kappa') \del_t \phi_3^{lm} - \kappa \del_r \phi_3^{lm} &= -(3\kappa' - 4 \mu) \phi_3^{lm} -\mu \alpha_2 \phi_4^{lm} ,\\
(1+t\kappa') \del_t \phi_2^{lm} - \kappa \del_r \phi_2^{lm}  &=-(3\kappa' - 3 \mu) \phi_2^{lm}   -\mu \alpha_0 \phi_3^{lm} ,\\
(1+t\kappa') \del_t \phi_1^{lm} - \kappa \del_r \phi_1^{lm} &= -(3\kappa' - 2 \mu) \phi_1^{lm}  -\mu \alpha_0 \phi_2^{lm} ,\\
(1+t\kappa') \del_t \phi_0^{lm} - \kappa \del_r \phi_0^{lm} &= -(3\kappa' - \mu) \phi_0^{lm}   -\mu \alpha_2 \phi_1^{lm} .
\end{aligned}
\end{equation}
We have defined the two quantities $\alpha_0=\sqrt{l(l+1)}$ and $\alpha_2=\sqrt{l(l+1)-2}$. We also abuse notation by omitting the superscript $(l,m)$ at the spinor components. Thus, we have to keep in mind that we get a system like the above for every admissible pair~$(l,m)$. Note, however, that the for the cases $l=0$ and $l=1$ the system looks different because $\alpha_2=0$ or $\alpha_0=0$.
These are advection equations along the null-directions defined by $l^a$ and $n^a$, from which can deduce a system of evolution equations and constraint equations.

\section{Analytical properties}
\label{sec:analyt-prop}

Before we describe the numerical implementation we want to briefly discuss some of the relevant analytical properties of this system. 

\subsection{Evolution and constraint systems}
\label{sec:evol-constr-syst}

We first split the system into evolution equations and constraints by taking appropriate linear combinations of the equations. This yields the evolution system
\begin{equation}
  \label{eq:11}
  \begin{aligned}
(1+t\kappa') \del_t \phi_0 - \kappa \del_r \phi_0 &= -(3\kappa' - \mu) \phi_0   -\mu \alpha_2 \phi_1 ,\\
\del_t \phi_1  &=  -  \mu \phi_1 +  \frac12 \mu \alpha_2 \phi_0 - \frac12 \mu \alpha_0 \phi_2,\\
\del_t \phi_2  &=    \frac12 \mu \alpha_0 \phi_1 - \frac12 \mu \alpha_0 \phi_3,\\
\del_t \phi_3  &=   \mu \phi_3 +  \frac12 \mu \alpha_0 \phi_2 - \frac12 \mu \alpha_2 \phi_4,\\
 (1-t\kappa') \del_t \phi_4 + \kappa \del_r \phi_4 &=  (3\kappa' -  \mu) \phi_4 +  \mu \alpha_2 \phi_3
  \end{aligned}
\end{equation}
and the constraints
\begin{equation}
  \label{eq:12}
  \begin{aligned}
     C_1 &\equiv - 2 \kappa \del_r \phi_1 + 6 r \mu' \phi_1 - 2 t \kappa' \mu \phi_1 +  \alpha_0 \mu (1-t\kappa') \phi_2 + \alpha_2 \mu (1+t\kappa') \phi_0 = 0,\\
  C_2 &\equiv -2 \kappa \del_r \phi_2 + 6 r \mu' \phi_2  + \alpha_0 \mu (1-t\kappa') \phi_3 + \alpha_0 \mu (1+t\kappa') \phi_1 = 0,\\
  C_3 &\equiv - 2 \kappa \del_r \phi_3 +  6 r \mu' \phi_3 + 2 t \kappa' \mu \phi_3 + \alpha_0 \mu (1+t\kappa') \phi_2 + \alpha_2 \mu (1-t\kappa') \phi_4 = 0.
  \end{aligned}
\end{equation}
Obviously, the evolution equations can be written in the form 
\[
\mathbf{A}^0 \del_t \Phi + \mathbf{A}^1 \del_r \Phi = \mathbf{B} \Phi
\]
with symmetric matrices $\mathbf{A}^0$ and $\mathbf{A}^1$. Furthermore, the matrix $\mathbf{A}^0 = \mathrm{diag}(1+t\kappa',1,1,1,1-t\kappa')$ is positive whenever $|t| < 1/|\kappa'|$. Thus, the evolution equations are symmetric hyperbolic on the domain, where $t$ is confined to these values. It is also straightforward to show that the constraints are propagated by the evolution equations, i.e., they propagate solutions of the constraints into solutions of the constraints: taking time derivatives of the constraint quantities $C_k$, using the evolution equations and the constraints we obtain the `subsidiary' system for the propagation of the constraint quantities
\begin{equation}
  \label{eq:14}
  \begin{aligned}
    \del_t C_1 &= - \mu C_1 - \frac12 \alpha_0 \mu C_2,\\
    \del_t C_2 &= \frac12 \alpha_0 \mu (C_1 - C_3),\\
    \del_t C_3 &= \mu C_3 + \frac12 \alpha_0 \mu C_2.
  \end{aligned}
\end{equation}
Since this is a homogeneous system of ODE, the claim follows immediately.

The matrix on the right hand side of this equation has eigen values $\lambda = 0, \pm\mu \sqrt{4-2l(l+1)}$. For $l\ge 2$ they all lie on the imaginary axis of the complex plane. So constraint violating modes due to numerical errors should not grow exponentially.

\subsection{Characteristics}
\label{sec:characteristics}

\begin{figure}[htb]
  \centering
  \includegraphics[width=0.7\textwidth]{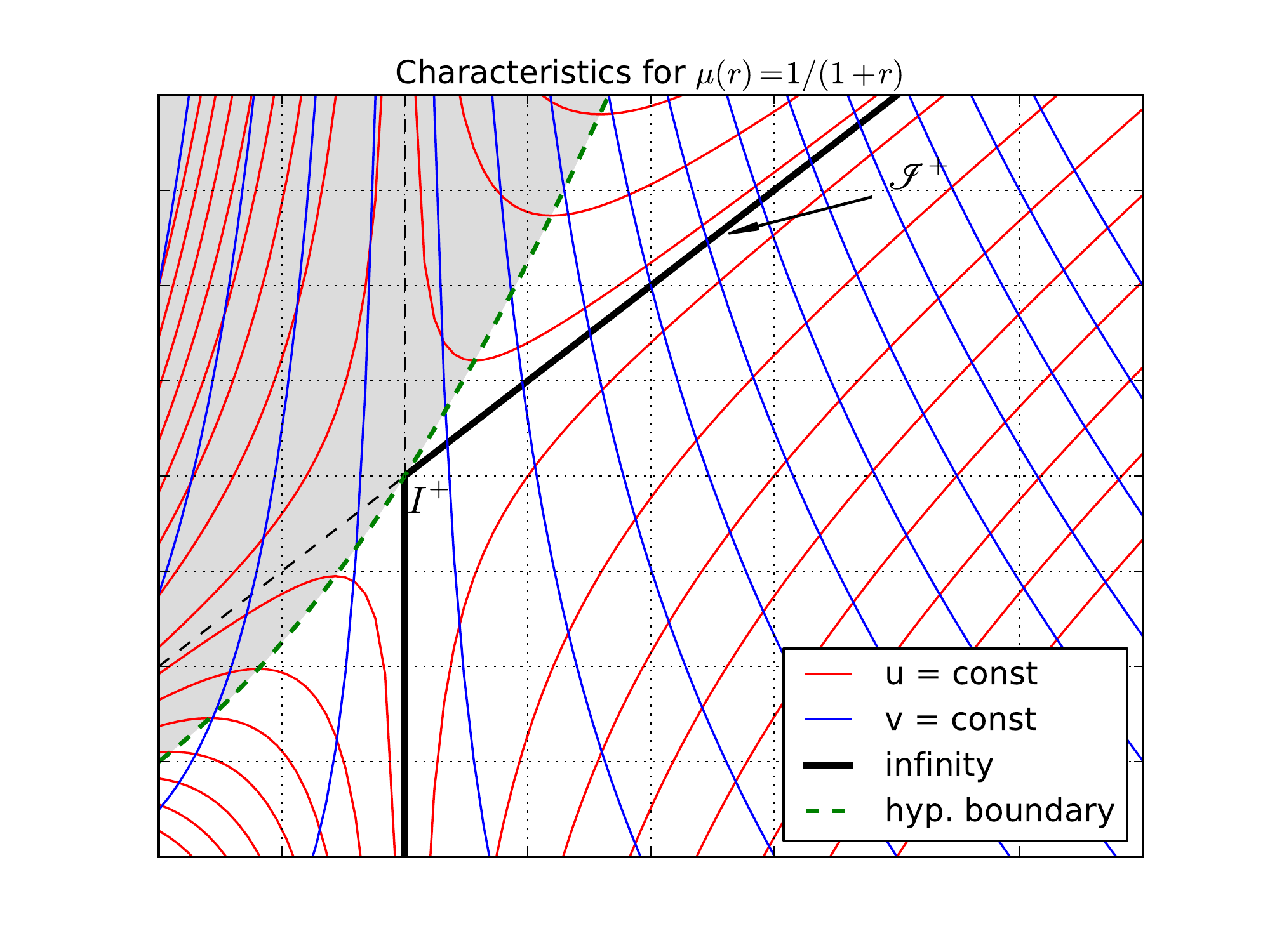}
  \caption{The characteristics of the evolution equations in a neighbourhood of 
$I^+$. The shaded region is the domain where the equations fail to be hyperbolic, i.e., where $t<\kappa'(r)$. The corresponding 
neighbourhood of $I^-$ is obtained by reflection at the lower border 
of the diagram accompanied by an interchange of $u$ and $v$. The thick (green) broken line indicates the boundary of the domain of hyperbolicity (shaded region).}
  \label{fig:characteristics}
\end{figure}

It is instructive to study the behaviour of the characteristics for the evolution 
system. Clearly, we have three different characteristics, the integral curves 
of the vectors $\del_t$, $l^a\del_a \propto (1-t\kappa') \del_t + \kappa \del_r$
and $n^a\del_a \propto (1+t\kappa') \del_t - \kappa \del_r$. It is easy to see 
that the latter two are exactly the lines of constant $u$ and $v$ defined in 
Eq.~\ref{eq:8}, respectively, and hence, they are null. 
In~Fig.~\ref{fig:characteristics} we show again the neighbourhood of 
$I$ and $\scri^+$ for $\mu(r)=1/(1+r)$. This time we show the characteristics 
also in the unphysical part of the diagram.
The non-shaded region bounded partly by the thick broken line is the domain of 
hyperbolicity of the evolution equations, i.e., the domain where $t<\pm |\kappa'(r)|$. Notice that every neighbourhood of 
$I^+$ contains regions where the hyperbolicity breaks down. Apart from the fact 
that one cannot hope to get existence and uniqueness of solutions beyond that 
region its presence also makes the numerical evolution challenging. For 
instance, setting up an initial boundary value problem with the left boundary 
at negative values of $r$ does not make sense if the evolution is to reach up 
to $I^+$. Even with the left boundary on $I$ it is not possible to go beyond 
$I^+$ since the evolution hits the non-hyperbolicity region.

\subsection{The constraint equations}
\label{sec:constraint-equations}

The constraint equations on $t=0$ are obtained from~\eqref{eq:12}. Define $\psi_k = \mu^{-3}\phi_k$ and let $l\ge2$ so that $\alpha_0$ and $\alpha_2$ both do not vanish. Then
\begin{equation}
  \label{eq:13}
    \begin{aligned}
      2 r\del_r \psi_1 &= \alpha_0 \psi_2 + \alpha_2 \psi_0,\\
      2 r \del_r \psi_2 &= \alpha_0 \psi_1 + \alpha_0 \psi_3,\\
      2 r \del_r \psi_3 &= \alpha_0 \psi_2 + \alpha_2 \psi_4.
  \end{aligned}
\end{equation}
Solutions of this system can be obtained by specifying functions $\psi_0(r)$ and $\psi_4(r)$, giving initial conditions at $r=0$ and then solving the equations for the remaining components. However,
for solutions with bounded derivatives at $r=0$ we cannot specify these data arbitrarily. Instead, for $r=0$ we have the equations
\begin{equation}
  \label{eq:r0}
    \begin{aligned}
      0 &= \alpha_0 \psi_2(0) + \alpha_2 \psi_0(0),\\
      0 &= \alpha_0 \psi_1(0) + \alpha_0 \psi_3(0),\\
      0 &= \alpha_0 \psi_2(0) + \alpha_2 \psi_4(0).
  \end{aligned}
\end{equation}
For $l\ge2$ this is an inhomogeneous linear system for the components $\psi_1$, $\psi_2$, $\psi_3$ with inhomogeneity given by $\psi_0$ and $\psi_4$. Since the homogeneous system has a non-trivial solution, we can obtain solutions of the inhomogeneous system only when $\psi_0(0)=\psi_4(0)$ holds. Then there is a 2-dimensional set of initial conditions for bounded solutions given by
\begin{equation}
\psi_0(0) = a,\quad \psi_1(0) = b ,\quad \psi_2(0) = - \frac{\alpha_2}{\alpha_0} a, \quad
\psi_3(0) = -b,\quad \psi_4(0) = a,\label{eq:16}
\end{equation}
where $a$ and $b$ are arbitrary complex constants. If $l=1$ then $\psi_0$ and $\psi_4$ vanish, since they have spin-weight $\pm2$, and this implies $\psi_2(0) = 0$. For $l=0$ the spin-weighted functions $\psi_1$ and $\psi_3$ vanish and $\alpha_0 = 0$. Then \eqref{eq:r0} does not impose any condition on $\psi_2(0)$.

The procedure to obtain initial data just described is suggested on physical grounds. One specifies the `wave-like' degrees of freedom represented by $\psi_0$ and $\psi_4$ and then computes from them the other components, which give rise to the energy-momentum. There is also another way to obtain initial data for the spin-2 system: we specify $\psi_2$ freely. When $l\ge1$ we use (\ref{eq:13},2) to obtain $\psi_1+\psi_3$ and we can specify the difference $\psi_1-\psi_3$ freely. Then, when $l\ge2$, we can use (\ref{eq:13},1,3) to obtain $\psi_0$ and $\psi_4$. Proceeding in this way the regularity conditions at $r=0$ are automatically satisfied. It has the advantage that no differential equations must be solved.

\subsection{The total characteristic $I$}
\label{sec:total-char-i}

We go back to the full system of equations~\eqref{eq:9} and for the discussion in this section we specialise to the `horizontal' conformal representation defined by $\mu(r)=1$. Evaluating the equations on $I$, where $\kappa(r)=0$ yields a system of eight PDE, which contain no radial derivatives, i.e., they are intrinsic to the cylinder. Therefore, no information can flow into and out of this set---it is a total characteristic. The evolution  of the spin-2 field on $I$ is entirely fixed by the initial data given at some initial time, say $t=0$. The intrinsic equations `transport' the initial data up (and down) to the location where null-infinity meets the cylinder. 

It was proven in~\cite{Friedrich:1998tc} 
that under the assumption of analyticity of the rescaled `unphysical' metric $g$ near 
space-like infinity the components of the spin-2 zero-rest-mass field admit an expansion of the form
\begin{equation}
 \label{eq:TE1} 
  \phi_k(t,r,\mathbf{e}) = \sum_{p=|s|}^\infty \frac1{p!}\chi_{k,p}(t,\mathbf{e})\,r^p,
\qquad \text{with } s = 2-k
\end{equation}
with\footnote{Friedrich uses a slightly different class of functions for this expansion. Here, we use the closely related well-known spin-spherical harmonics. See app.~\ref{sec:spin-spher-harm} for a detailed discussion of the relationship between both systems of functions.}
\begin{equation}
 \label{eq:TE2} 
\chi_{k,p}(t,\mathbf{e})=\sum_{l=|s|}^p\sum_{m=-l}^{l}\phi_{k,p}^{lm}(t)\,\Y{s}{lm}(\mathbf{e}),
\end{equation}
where $\chi_{k,p}(t,\mathbf{e})=\partial_r^{(p)}\phi_k(t,r,\mathbf{e})|_{r=0}$. Note, that the number of $l$-modes is bounded by the differentiation order $p$.

The analytical behaviour of the solutions of these `transport equations' near the cylinder is completely known, see~\cite{Friedrich:2003hp,ValienteKroon:2002dx}, so we will only briefly summarise the analysis here. After decomposing the components into spherical modes we obtain the expansions
\begin{equation}
  \label{eq:18}
  \phi_k^{lm}(t,r) = \sum_{p=|s|}^\infty \frac1{p!}\phi_{k,p}^{lm}(t)\,r^p,
\qquad \text{with } s = 2-k  
\end{equation}
with $\phi_{k,p}^{lm}(t)=\partial_r^{(p)}\phi_k^{lm}(t,r)|_{r=0}$ and $l\le m$. 

Let $p\ge2$ since the cases $p=0,1$ are special. Taking radial derivatives of the equations~\eqref{eq:fullminus} and \eqref{eq:fullplus} and then restricting to $r=0$ we can derive coupled second order equations for the coefficient functions $\phi_{k,p}^{lm}(t)$ of the five components of the spin-2 field (where a dot denotes differentiation with respect to $t$)
\begin{equation}
    (1-t^2)\, \ddot \phi_{k,p}^{lm} + \left[2s + 2(p-1)t \right]\, \dot \phi_{k,p}^{lm} + [l(l+1) - p(p-1)] \,\phi_{k,p}^{lm} = 0, \qquad k=0:4,\label{eq:ddotPhik}
\end{equation}
which is a Jacobi differential equation. From the discussion in~\cite{Szego:1959tn} we find that this equation has regular polynomial solutions if and only if $|s|\le l < p$. On the other hand, from~\eqref{eq:TE1} we see that $l=p$ is an admissible term in the expansion so that the highest $l$-terms in this expansion cannot be regular. In fact, one can show (see~\cite{Szego:1959tn}) that they contain logarithmic divergences at $t=\pm1$. 

The solutions for $l=p$ are of the form
\begin{equation}
  \label{eq:19}
  \begin{aligned}
    \phi_{kp}^{pm}(t) = A^m_{kp}\,&P_{2p}^{-s-p,s-p}(t) \\
    &+B^m_{kp}\;\frac2{1-t}\;
    {}_2F_1(1,1+s+p;2p+2;\frac2{1-t})
  \end{aligned}
\end{equation}
for some constants $A^m_{kp}$ and $B^m_{kp}$. The functions $P_{2p}^{-s-p,s-p}(t)$ are Jacobi polynomials, so they are regular, while  the hypergeometric functions ${}_2F_1(1,1+s+p;2p+2;\frac2{1-t})$ contain the logarithmic divergences.

To simplify the notation we will drop the indices $m$ and $p$ from these coefficients because they are irrelevant to the following discussion. Inserting the solutions~\eqref{eq:19} into the eight first order equations~\eqref{eq:fullminus}, \eqref{eq:fullplus} shows that the coefficients $A_{k}$ resp. $B_{k}$ are linearly related so that we can determine them all given only one of them, say $A:=A_{2}$ resp. $B:=B_{2}$:
\begin{equation}
A_{0} = A_4 = \frac{\alpha_0\alpha_2}{(p+1)(p+2)}A,\quad 
A_{1} = A_3 = -\frac{\alpha_0}{p+1}A,\label{eq:15}
\end{equation}
and
\begin{equation}
B_{0} = B_4 = \frac{\alpha_0\alpha_2}{p(p-1)}B,\quad 
B_{1} = B_3 = \frac{\alpha_0}{p}B.\label{eq:20}
\end{equation}
This implies that we can eliminate the logarithmic terms by putting $B=0$. We can express this condition in terms of the initial conditions at $t=0$ as follows: let
\begin{equation}
\underline{\phi}_k=\phi^{pm}_{kp}(0)=A_k 2^{-2p} + B_k \underline{f}_k,\qquad k=0:4\label{eq:21}
\end{equation}
where $\underline{f}_k = {}_2F_1(1,1+s+p;2p+2;2)$. By using~\eqref{eq:15} and eliminating $A$ between successive equations~\eqref{eq:21} we obtain the four relations
\begin{equation}
  \label{eq:22}
  \begin{aligned}
    \alpha_2\;\underline{\phi}_1 + (p+2)\underline{\phi}_0 &=
    \left(\underline{f}_1 + \frac{p+2}{p-1} \underline{f}_0\right)
    \frac{\alpha_2 \alpha_0}p B\\
    \alpha_0\;\underline{\phi}_2 + (p+1)\underline{\phi}_1 &=
    \left(\underline{f}_2 + \frac{p+1}{p} \underline{f}_1\right)
    \frac{\alpha_0}p B\\
    \alpha_0\;\underline{\phi}_2 + (p+1)\underline{\phi}_3 &=
    \left(\underline{f}_2 + \frac{p+1}{p} \underline{f}_3\right)
    \frac{\alpha_0}p B\\
    \alpha_2\;\underline{\phi}_3 + (p+2)\underline{\phi}_4 &=
    \left(\underline{f}_3 + \frac{p+2}{p-1} \underline{f}_4\right)
    \frac{\alpha_2 \alpha_0}p B\\
  \end{aligned}
\end{equation}
Choosing initial data such that the left hand side in any of these relations vanishes implies that all left hand sides in~\eqref{eq:22} vanish (since one can show that the terms in parentheses on the right hand sides never vanish) and that the solution for $\phi^{pm}_{kp}(t)$ on the cylinder has no logarithmic divergences.

Notice, that these relations hold for $p\ge2$. Since one can show~\cite{Friedrich:1998tc} that the coefficients of $r^0$ and $r^1$ in~\eqref{eq:18} are necessarily free of logarithms the first appearance of logarithmic divergences can occur for $p=2$. We will use this fact in sect.~\ref{sec:transport-eqs}.

By going back to the expansion of the spinor components one can interpret~\eqref{eq:22} as differential equations on the initial hyper-surface $t=0$, relating angular and radial derivatives of the spinor components evaluated at $r=0$. It is a consequence of a truly remarkable result of Friedrich's~\cite{Friedrich:1998tc} which shows that these relations are the components of the equations 
\begin{equation}
  \label{eq:23}
  D_{(A_1B_1}D_{A_2B_2}\ldots D_{A_pB_p}B_{CDEF)} = 0
\end{equation}
evaluated on the cylinder. Here, $D_{AB}$ denotes the intrinsic (space spinor) derivative operator on the initial hyper-surface induced by the metric $g$ and $B_{ABCD}$ is the linearised Cotton spinor, expressible in our context as 
\[
B_{ABCD}= 2 D_{(A}{}^E \Omega\; \phi_{BCD)E} + \Omega\, D_{(A}{}^E \phi_{BCD)E} ,
\] 
where $\Omega$ is the conformal factor relating the flat Minkowski space metric to the metric~\eqref{eq:5}.

\subsection{Boundaries and boundary conditions}
\label{sec:bound-bound-cond}
In this paper we discuss the solution of the spin-2 equations in a neighbourhood of the cylinder at $r=0$ for times $t\le1$. Having decomposed the equations into different modes due to the rotational symmetry we have a $1+1$ dimensional system. Our spatial domain is the interval $[0,1]$ and we are interested in the time development from $t=0$ up to as close to $t=1$ as possible.

We need to discuss the boundaries at $r=0$ and at $r=1$. Since in this work we are only interested in the behaviour of the solution near space-like infinity, i.e., near $r=0$, we introduce an artificial boundary at $r=1$. A more adequate treatment would be based on the conformal compactification of Minkowski space to a part of the Einstein cylinder and then have a radial interval with space-like infinity at one end and the centre of the polar coordinate system at the other. However, then we would have the usual problems with coordinate singularities, which we will address in future work.

At this artificial `outer' boundary at $r=1$ there is one ingoing characteristic, corresponding to $\phi_0$ and an outgoing characteristic corresponding to $\phi_4$. So we need to specify $\phi_0$ at $r=1$. All the other characteristics are tangent to the boundary. Since the subsidiary system~\eqref{eq:14} has no ingoing modes at $r=1$ there are no further conditions on the specification of $\phi_0$ apart from corner conditions at $(t,r)=(0,1)$.

The other boundary, $r=0$, is a total characteristic, i.e., there are no ingoing or outgoing modes. All characteristic directions are tangent to the boundary. Therefore, we do not need to do anything special there. 

Let us point out that it would be possible to put an artificial boundary at some location $r<0$, where again there would be one ingoing and one outgoing mode, i.e., one boundary condition. However, as discussed in sect.~\ref{sec:characteristics} it would be impossible with this setup to reach $t=1$ because of the breakdown of hyperbolicity in any neighbourhood near $t=1$.

Also, note that the `outer' boundary at $r=1$ (which is really the inner boundary from the point of view of the original Minkowski space picture) is an artificial boundary which exists only due to our conformal representation of the neighbourhood of spatial infinity based on an involution of Minkowski space. If we had chosen a different representation such as the one based on the standard representation of Minkowski space in the Einstein static universe, then we could have eliminated this artificial boundary entirely and we could have studied the system globally on one computational domain. This will be discussed in more detail elsewhere.

\section{Numerical implementation and results}
\label{sec:numer-impl}

\subsection{Numerical algorithms}
\label{sec:numer-routines}

We use the method of lines to solve the equations numerically. Each space-like surface, $\{(t,r):r\in[0,1]\}$, is represented as a grid with constant spacing $\Delta r$. The spatial discretisation is obtained by replacing the spatial derivatives with a summation by parts (SBP) operator, \cite{Strand:1994ef,Carpenter1994TimeStable}. SBP operators are finite difference approximations to derivatives that obey a discrete version of integration by parts, \cite{Strand:1994ef}. This ensures that the matrix, that represents the first order spatial derivative, is semi-bounded with respect to a suitable norm on a vector space, \cite{Strand:1994ef}, and it allows for energy estimates to be calculated for our discretization of our system of equations, \eqref{eq:11}, by following the arguments used in the continuous case.

We have used the minimum bandwidth, restricted full norm, fourth order (on the interior and third order on the boundary) SBP operator given in \cite[page 65]{Strand:1994ef}. We have tried several other SBP operators from the papers \cite{Strand:1994ef,Diener2007Optimized,Carpenter1999Stable} but have found that our choice gives the smallest absolute error over each grid point.

The semi-discrete system thus obtained is then solved by the standard 4th order Runge-Kutta solver. The time-step $\Delta t$ used in the RK4 method is computed either with a constant CFL factor $C = \Delta t/\Delta r$ or by computing the maximal possible time-step from the information about the characteristic speeds available at each instant of time using the criterion that the numerical domain of dependence be larger than the analytical domain of dependence. This `adaptive' time-step was particularly useful for runs of the cases with horizontal $\scri$ ($n=0$, see sect.~\ref{sec:minkowski-space-time}) because in this case one of the characteristic speeds diverge for $t\to 1$. This has the consequence that any choice of constant CFL factor eventually leads to instability of the code near $r=1$. The adaptive time-step allows us to approach $t=1$ arbitrarily closely. Of course, the draw-back is that the time-step becomes arbitrarily small so that we cannot reach $t=1$ exactly.

The initial data are either specified by hand from exact solutions or by solving the constraint equations using one of the two methods discussed in sect.~\ref{sec:constraint-equations}. When solving the ODE's we use the DOP853 method, a Runge-Kutta method of order 8(5,3) due to Dormand and Prince as described in~\cite{Hairer:1993vh}. 

In line with the recommendation of~\cite{Mattsson:2003dd} we implement the boundary conditions 
by using the simultaneous approximation term (SAT) method (see~\cite{Carpenter1994TimeStable,Carpenter1999Stable}), a penalty method for enforcing boundary conditions. The enforcing equation, in semi-discrete notation, is
\begin{equation}
  \begin{aligned}
    \dot\phi_{0,i} = \frac{1}{1+t\,\kappa'_i}\biggl(\kappa_i\, (\mathbf{Q}\phi_0)_i - &(3 \kappa'_i - \mu_i)\,\phi_{0,i} - \mu_i\alpha_2\phi_{1,i}\biggr)\\
    &- \tau \frac{\kappa_N}{1+t\kappa'_N} \delta_{i,N} (\phi_{0,N}-b_r(t)),
  \end{aligned}
\qquad i=1:N
\end{equation}
where $\phi_0$, $\phi_1$, $\kappa$, $\kappa'$ and $\mu$ are grid-functions, i.e., vectors of length $N$ with $\phi_{0,i}$ etc.\ being the $i$'th component. $\mathbf{Q}$ is an SBP operator approximating the first derivative with respect to $r$ represented as a constant $N\times N$ matrix acting on $\mathbb{R}^N$ and $b_r(t)$ is the free function given on the right boundary. The parameter $\tau$ is suitably adjusted to enforce the boundary conditions in a stable way. For details we refer to~\cite{Strand:1994ef,Carpenter1999Stable}. 

We have implemented the numerical algorithms discussed above in a Python code, which relies on the numerical routines of the NumPy and SciPy packages.

\subsection{Code test with an exact solution}
\label{sec:code-test-exact}

\begin{figure}[htb]
  \centering
  \includegraphics[width=0.8\linewidth]{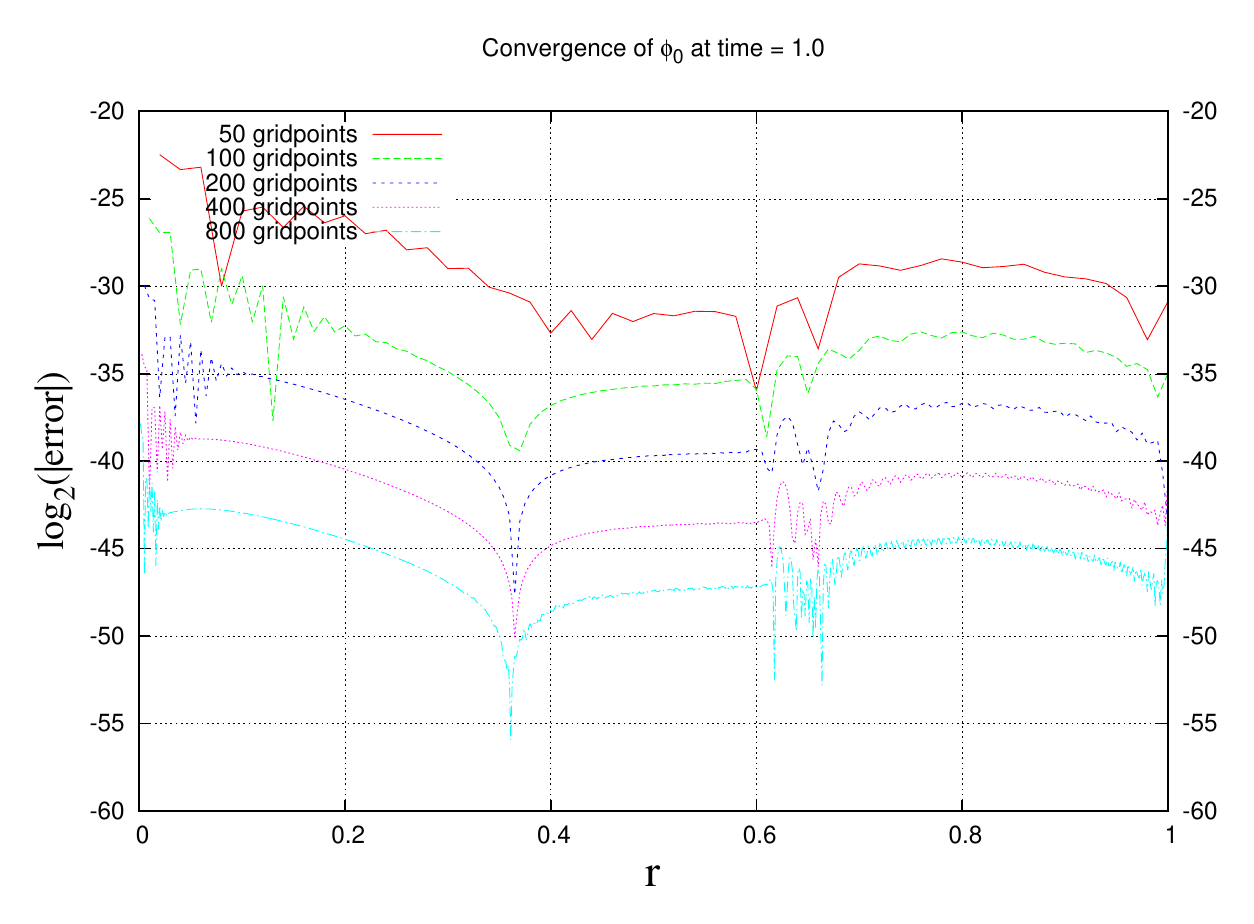}
  \caption{Convergence of $\phi_0$ to the exact solution at time $t=1$. }
  \label{fig:conv_exact_plots_0}
\end{figure}

\begin{figure}[htb]
  \centering
  \includegraphics[width=0.8\linewidth]{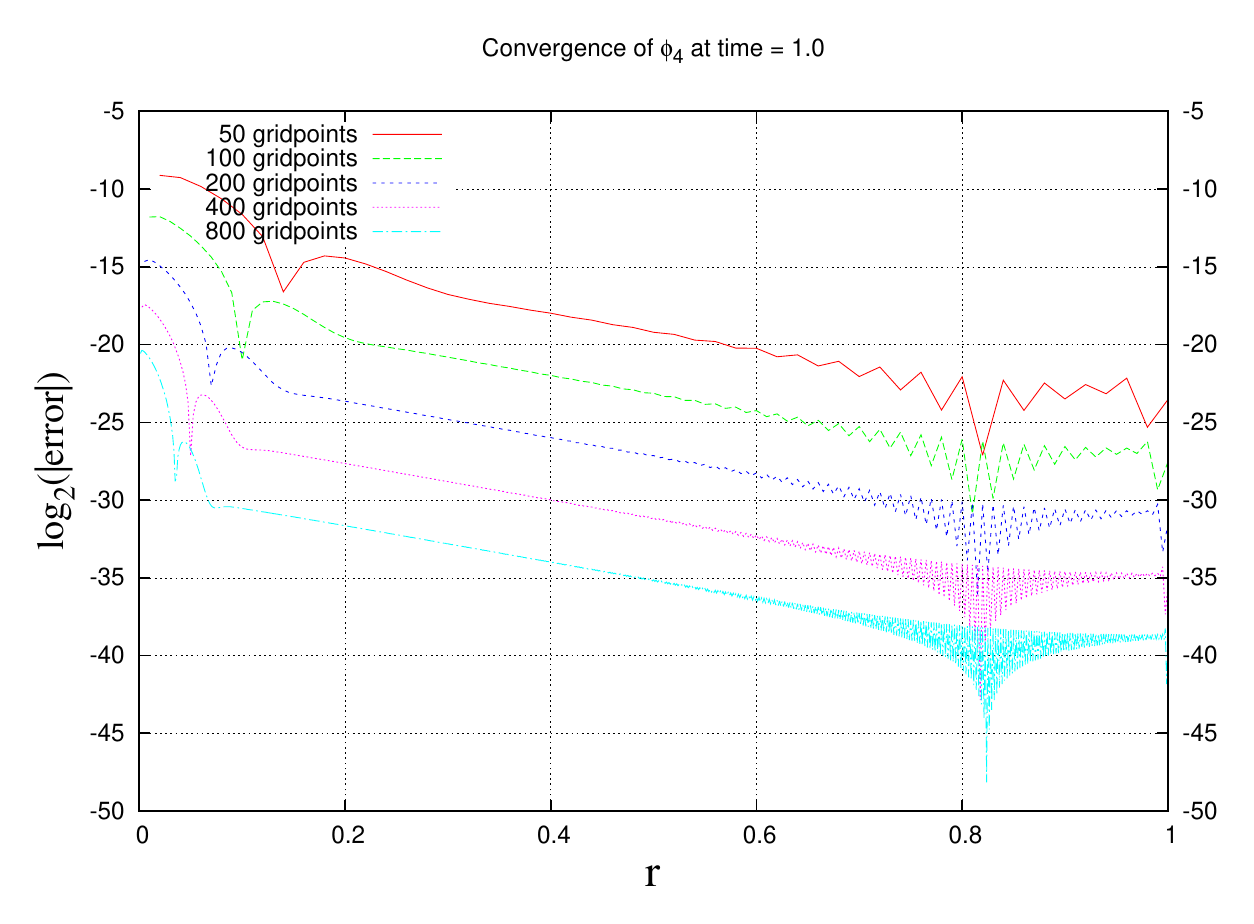}
  \caption{Convergence of $\phi_4$ to the exact solution at time $t=1$. }
  \label{fig:conv_exact_plots_4}
\end{figure}

Here we will reproduce numerically the exact solution of the system \eqref{eq:11}-\eqref{eq:12} 
derived in appendix~\ref{sec:an-exact-solution}. This solution refers to the case with diagonal $\scri$ ($n=1$ see sect.~\ref{sec:minkowski-space-time}). For this we prescribe initial conditions 
of the form (set $t=0$ in \eqref{eq:A13})
\begin{equation}
 \label{eq:ES1}
    \phi_0 = \frac{r^2}{(1+r)^3}, 
    \phi_1 = \frac{2 r^2}{(1 + r)^3},
    \phi_2 = \frac{\sqrt{6} r^2}{(1+ r)^3},
    \phi_3 = \frac{2 r^2}{(1+ r)^3}, 
    \phi_4 = \frac{r^2}{(1+ r)^3}.    
\end{equation}
We also impose, as discussed above, only a boundary condition for $\phi_0$ at $r=1$:
\begin{equation}
    \phi_0(t, 1) = \frac{(2 - t)^4}{128}.    
\end{equation}
Evolving the initial data~\eqref{eq:ES1} with a fixed time-step and computing the difference between the computed and the exact solution we obtain the convergence plots for the components $\phi_0$, 
$\phi_4$ at time $t=1$ as presented in Figures~\ref{fig:conv_exact_plots_0} and~\ref{fig:conv_exact_plots_4}.  

Table~\ref{tab:conv_exact} lists the $\log_2$ of the absolute error in the $L^2$-norm and the convergence rates at time $t=1$ for $\phi_0, \phi_4$.
\begin{table}[htb]
  \begin{tabular}{c||rr|rr}
      & $\log_2(||\Delta||_2)$ & Rate & $\log_2(||\Delta||_2)$ & Rate \\\hline\hline
    50  & -24.899275 &        &  -11.322955 &         \\
    100 & -29.057909 & 4.1586 &  -14.245253 & 2.9223 \\ 
    200 & -33.378371 & 4.3205 &  -17.326037 & 3.0808 \\ 
    400 & -37.781307 & 4.4029 &  -20.487582 & 3.1615 \\
    800 & -42.216813 & 4.4355 &  -23.689231 & 3.2016
  \end{tabular}
  \caption{Absolute error $\Delta$ compared to the exact solution using the $L^2$ norm and convergence rates at time $t=1$ for $\phi_0, \phi_4$. The calculation was done with a fixed time-step. The first two columns refer to $\phi_0$, the others to $\phi_4$.}
  \label{tab:conv_exact}
\end{table}
We find the expected overall 4th order convergence. However, note that this refers to the time $t=1$, i.e., when the equation become singular on the left boundary. So we can reach $t=1$ without loss of convergence. Running beyond $t=1$, however, very quickly leads to instabilities and code crash, as it is to be expected from the loss of hyperbolicity of the equations near $t=1$ on the left boundary.

The convergence plots for $\phi_0$ (Fig.~\ref{fig:conv_exact_plots_0}) show that the inflow  boundary at $r=1$ is very well behaved, similarly the total characteristic at $r=0$ shows no problems in the case of $\phi_4$, which propagates away from it towards increasing $r$. We do, however, see high frequency features at $r=0$ for $\phi_0$ and at $r=1$ for $\phi_4$. In both cases these are numerical artifacts. They indicate that the modes hitting the boundary from inside the grid are reflected due to numerical inaccuracies. We could eliminate them both by using small amounts of artificial dissipation but have chosen not to in the present work because, as indicated at the beginning of sect.~\ref{sec:bound-bound-cond}, the `outer' boundary is an artificial boundary which we intend to eliminate anyway.

\subsection{Code test with general initial data}
\label{sec:code-test-with}

We specify initial data using the algebraic method described in sect.~\ref{sec:constraint-equations}. We specify 
\[
\phi_2(0,r) = 
\begin{cases}
  (4\,r/b^2)^{16} (r-b)^{16} & 0\le r \le b\\
  0 & b\le r \le 1
\end{cases}.
\] 
This gives a `bump' centered at $r=b/2$ and we chose $b=4/5$ for these runs. We also chose $\phi_1(0,r)=\phi_3(0,r)$. Then all the spinor components are fixed by the constraint equations at $t=0$, see sect.~\ref{sec:constraint-equations}. They have compact support in the interval $0\le r \le 1$ and in order to satisfy the corner conditions we choose the boundary condition $\phi_0(t,1) = 0$. We run the tests with the two different conformal representations specified by $n=0$ and $n=1$. Tab.~\ref{tab:convergence_rates_0} and tab.~\ref{tab:convergence_rates_1} show the results for $l=2$. In both cases we obtain the expected 4th order convergence in $\phi_0$. However, in the `horizontal' representation we see a loss of convergence in the $\phi_4$ component, which is due to the fact that the characteristics for $\phi_4$ approach $\scri^+$ so that the characteristic speed becomes large. The decrease in the convergence rate might be due to the fact that the code is leaving its stability region.

\begin{table}[htb]
  \centering
\[  \begin{array}{c||rr|rr}
   &  \log_2(||\Delta||_2) &   \text{Rate} &   \log_2(||\Delta||_2) &  \text{Rate}    \\\hline\hline
50 &   1.383744   &&    6.860341 &           \\
100&   -2.646956 &  4.0307  &   2.268950 &  4.5914 \\
200&  -6.615024 &  3.9681  &  -2.951419 &  5.2204    \\
400&  -10.692111 &  4.0771  &  -5.655223 &  2.7038    \\ 
  \end{array}
\]
  \caption{Convergence rates of the absolute error $\Delta$ in the $L^2$ norm at $t=0.96$ for compactly supported initial data evolved in the `horizontal' conformal representation with $n=0$. The error is computed with respect to a higher resolution run with 800 grid points. The first two columns refer to $\phi_0$, the others to $\phi_4$.}
  \label{tab:convergence_rates_0}
\end{table}

\begin{table}[htb]
  \centering
\[  \begin{array}{c||rr|rr}
   &  \log_2(||\Delta||_2) &  \text{Rate} &   \log_2(||\Delta||_2) &  \text{Rate}   \\\hline\hline
50  &  1.447922     &&                  2.228200             &        \\
100 &  -2.508450    &  3.9564        &  -1.646880           & 3.8751 \\
200 &  -6.479996    &  3.9715        &  -5.645109           & 3.9982 \\
400 &  -10.558692   &  4.0787        &  -9.729219           & 4.0841
  \end{array}
\]
  \caption{Convergence rates of the absolute error $\Delta$ in the $L^2$ norm at $t=1.0$ for compactly supported initial data evolved in the `diagonal' conformal representation with $n=1$. The error is computed with respect to a higher resolution run with 800 grid points. The first two columns refer to $\phi_0$, the others to $\phi_4$.}
  \label{tab:convergence_rates_1}
\end{table}

Finally, in Tab.~\ref{tab:convergence_rates_10_1} we show the convergence rates for a higher mode ($l=10$) using  compactly supported data in the `diagonal' representation.

\begin{table}[htb]
  \centering
\[  \begin{array}{c||rr|rr}
   &  \log_2(||\Delta||_2) &  \text{Rate} &   \log_2(||\Delta||_2) &  \text{Rate}   \\\hline\hline
50  &  -3.822906 &&            -1.422306 &              \\
100 &  -7.773146 &  3.9502  &  -5.454578 &  4.0323     \\
200 &  -11.755335&  3.9822  &  -9.437786 &  3.9832     \\
400 &  -15.836066&  4.0807  &  -13.522334&  4.0845    
  \end{array}
\]
  \caption{Convergence rates of the absolute error $\Delta$ in the $L^2$ norm at $t=1.0$ for compactly supported initial data with $l=10$ evolved in the `diagonal' conformal representation with $n=1$. The error is computed with respect to a higher resolution run with 800 grid points. The first two columns refer to $\phi_0$, the others to $\phi_4$.}
  \label{tab:convergence_rates_10_1}
\end{table}

\subsection{Constraint violation}
\label{sec:constraint-violation}

\begin{figure}[htb]
  \centering
  \includegraphics[width=0.8\linewidth]{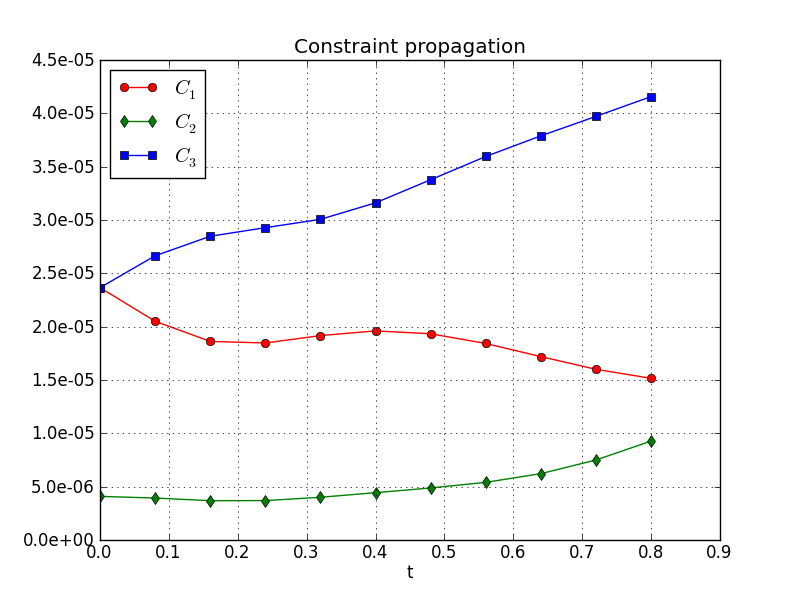}
  \caption{Constraint violation for several values of $t \in [0,0.8]$. We show the $L^2$-norm of the three constraint quantities.}
  \label{fig:constraintpropagation}
\end{figure}

Our next topic is the behaviour of the constraint quantities defined in~\eqref{eq:12}. We use the same initial and boundary data as in sect.~\ref{sec:code-test-with} with the initial bump centered at $r=0.3$. We take $l=2$ and $n=1$. At each time-step we evaluate the three constraint quantities. In Fig.~\ref{fig:constraintpropagation}
we show the $L^2$ norm of the three constraint quantities at selected instants of time in the interval $[0,0.8]$. The size of the constraint violation remains almost the same. This is clearly due to the fact that we are dealing with a linear system and also only for a short time interval. We have also checked the convergence of the constraint quantities with increasing resolution. We find the expected 4th order convergence to zero.

\subsection{Transport equations}
\label{sec:transport-eqs}

Now we are in a position to attempt a reproduction of Friedrich's analytical results on 
the cylinder for $\mu(r) = 1$. Recall from sect.~\ref{sec:total-char-i} that the components of the spin-2 zero-rest-mass field $\phi_{ABCD}$ admit an expansion of the form
\begin{equation}
  \phi_k(t,r,\mathbf{e}) = \sum_{p=|s|}^\infty \frac1{p!}\chi_{k,p}(t,\mathbf{e})\,r^p,
\qquad \text{with } s = 2-k
\end{equation}
with
\begin{equation}
\chi_{k,p}(t,\mathbf{e})=\sum_{l=|s|}^p\sum_{m=-l}^{l}\phi_{k,p}^{lm}(t)\,\Y{s}{lm}(\mathbf{e}),
\end{equation}
where $\chi_{k,p}=\partial_r^{(p)}\phi_k|_{r=0}$. The discussion in~\ref{sec:total-char-i} showed that the time dependent coefficients $\phi_{k,p}^{lm}(t)$ can contain logarithmic divergences at $t=\pm1$ unless one restricts their initial values at $t=0$. The singular behaviour affects those coefficients with $p=l$ and $p\ge2$.

In our numerical computation we have already decomposed the spinor components $\phi_k$ into their spherical harmonic pieces. So each function $\phi_k(t,r)$ in~\eqref{eq:fullminus} and~\eqref{eq:fullplus} already corresponds to a particular pair $(l,m)$ and each of these modes has an expansion near $r=0$ of the following type
\begin{equation}
  \label{eq:17}
  \phi_k(t,r) = \sum_{p=|s|}^\infty \sum_{l=|s|}^p \frac1{p!} \phi_{k,p}^{lm}(t) r^p, \qquad \text{with } s = 2-k.
\end{equation}
We choose the functions $\phi_0(0,r)$ and $\phi_4(0,r)$ at $t=0$ in such a way that 
\[
\phi_{0,2}^{2m} = \lim_{r\to0} \phi_0(0,r)/r^2 \ne0,
\] 
and similarly for $\phi_4(0,r)$. Specifically, we take
\begin{equation}
 \label{eq:TE4}
  \phi_0(0,r) = 8\,r^2(r-1)^{36}, \quad \phi_4(0,r) = -8\,r^2(r-1)^{36}.
\end{equation}
With these functions we solve the constraints at $t=0$ and then we evolve the system for $l=2$ and arbitrary $m$ using the conformal representation with $n=0$. We use the boundary condition $\phi_0(t,1) = 0$ at the `outer' boundary. The computation was done with the adaptive time-step in order to come as close to the singular time $t=1$ as possible. It was stopped at $t=0.9999$.

\begin{figure}[htb]
  \centering
  \includegraphics[width=0.8\linewidth]{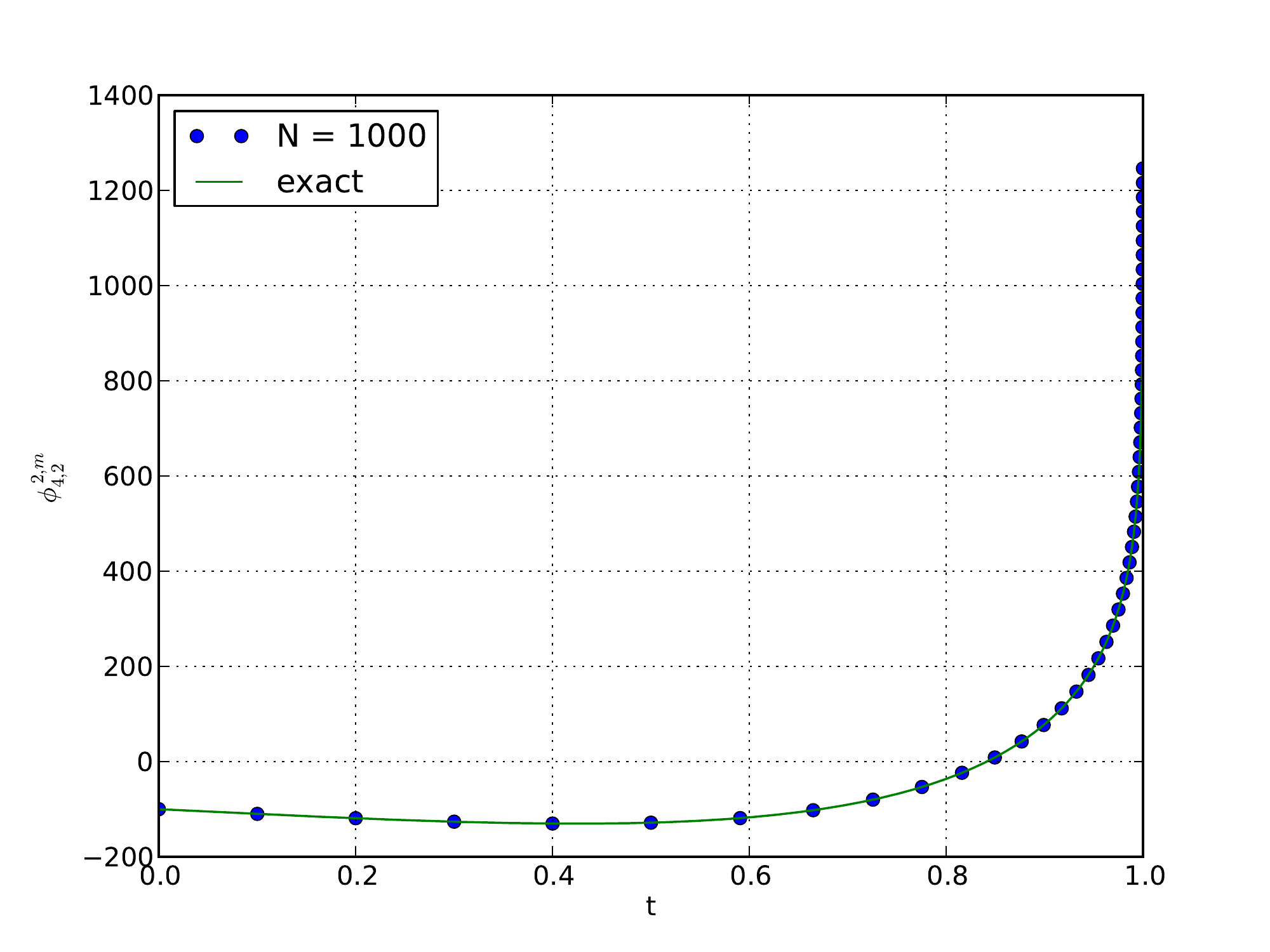}
  \caption{The first singular coefficient $\phi_{4,2}^{2,m}$ along $r=0$ for times $0\le t \le 1$.}
  \label{fig:a422m}
\end{figure}

\begin{figure}[htb]
  \centering
  \includegraphics[width=0.8\linewidth]{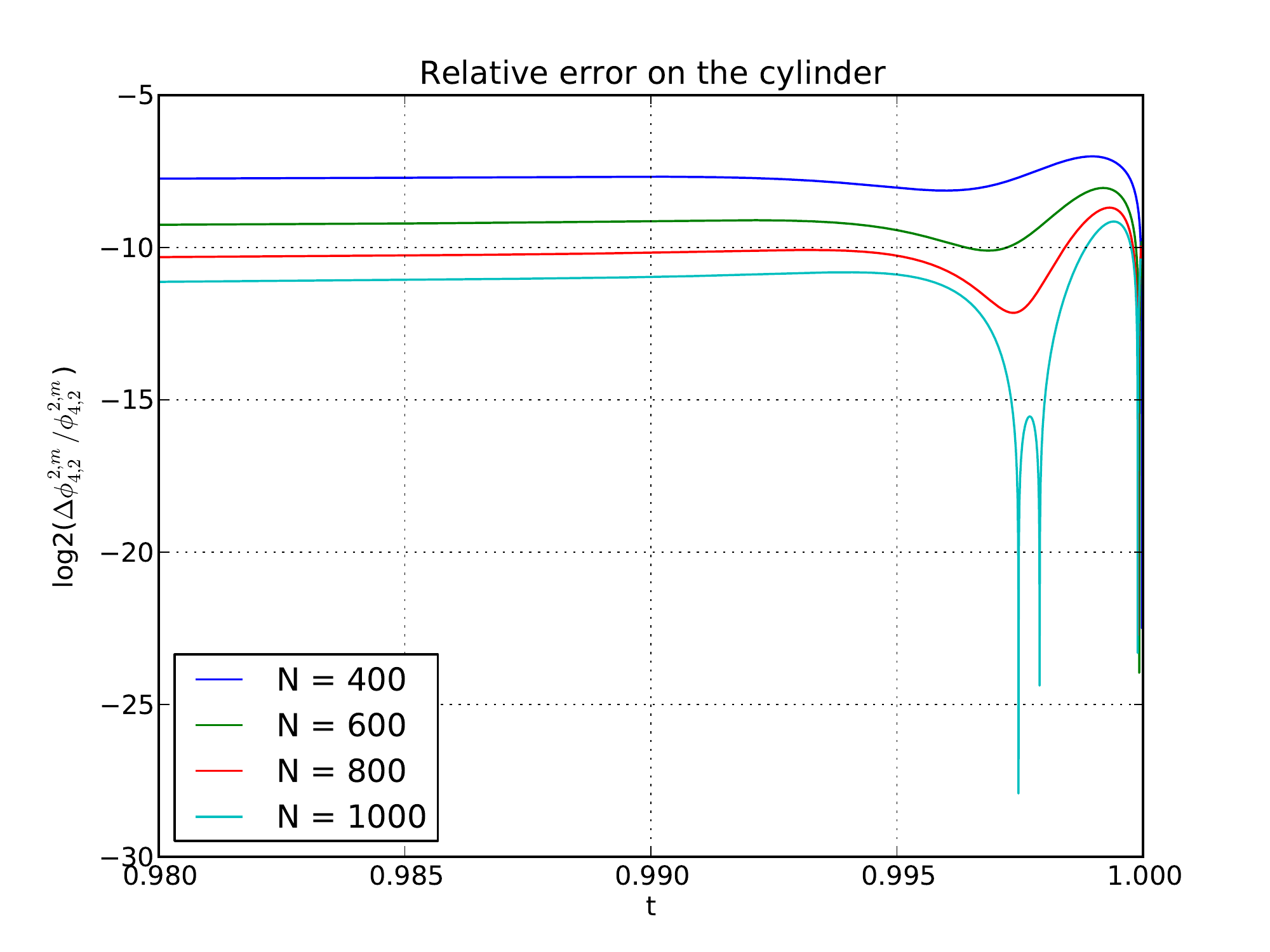}
  \caption{Relative error in the computation of the first singular coefficient $\phi_{4,2}^{2,m}$ along $r=0$ for times $0.98\le t \le 1$. }
  \label{fig:relative_error}
\end{figure}

On the other hand, following the procedure outlined in sect.~\ref{sec:total-char-i} we can compute the exact form of the singular coefficients~$\phi_{k,p}^{2,m}$ along the cylinder as a function of~$t$, given the initial data
\[
\phi_{k,2}^{2,m}(0) = \frac12\del_{rr}\phi_k(0,0).
\]
With the specified data we obtain the coefficients
\begin{equation}
  \begin{aligned}
    \phi_{0,2}^{2,m}(t) &= 16 - 19 t + 12 t^2 - 3 t^3 + \frac32 (1-t)^4 \ArcTanh(t), \\
    \phi_{1,2}^{2,m}(t) &= 2 (4 + t - 6 t^2 + 3 t^3 + 3 (1-t)^3 (1 + t) \ArcTanh(t)), \\
    \phi_{2,2}^{2,m}(t) &= \sqrt{6} (5 t - 3 t^3 + 3 (1-t)^2(1+t)^2 \ArcTanh(t)), \\
    \phi_{3,2}^{2,m}(t) &= 2(-4 + t + 6 t^2 + 3 t^3 + 3 (1-t) (1 + t)^3 \ArcTanh(t)), \\
    \phi_{4,2}^{2,m}(t) &= -16 - 19 t - 12 t^2 - 3 t^3 + \frac32 (1 + t)^4 \ArcTanh(t).
\label{eq:TE3}
  \end{aligned}
\end{equation}
In order to  check how the code can cope with the singular behaviour at $r=0$ we use the generated numerical solution in the computational domain with $0 \le t \le 1$. In view of the expansion of the field components $\phi_k$ near $r=0$ we compute at every time-step
\begin{equation}
 \label{eq:TE5}
\phi_{k,2}^{2,m}(t) =\frac{1}{2}\del_{rr}\phi_k(t,0).
\end{equation}
We evaluate the second derivative by using at least 4th order accurate one-sided finite difference operators which can be found in~\cite{Fornberg:1996to} and compare the result with the exact solution given in~\eqref{eq:TE3}. Fig.~\ref{fig:a422m}
shows the numerically computed coefficient $\phi_{4,2}^{2,m}$ and its exact value over the entire time interval $[0,1]$. In Fig.~\ref{fig:relative_error} we zoom in to the last stages of the evolution and present the relative error for times larger than $0.9$. We see that we get roughly the expected 4th order convergence even though the accuracy is not great. This is clearly due to the numerical differentiation in the radial direction to obtain the coefficient in front of $r^2$.

\section{Discussion}
\label{sec:discussion}

In this paper we have presented tests and preliminary applications for the linearised general conformal field equations in a setting near space-like infinity $i^0$ of Minkowski space-time. We have used Friedrich's conformal representation of a neighbourhood of $i^0$ which separates the `end points' of $\scri\pm$ by a cylinder~$I$.

Our motivation for this is the possibility to study entire space-times on a finite grid without the introduction of artificial boundaries. In contrast to traditional numerical approaches with a boundary at a finite distance from the source we are guaranteed that the corresponding solution is indeed asymptotically flat. In this setting we have access to asymptotic quantities such as the ADM mass and the radiation field at infinity. Furthermore, it is suggested by Friedrich's analytical results that, in principle, we have complete control over the smoothness of null infinity by choosing the behaviour of initial data on $I^0$. This issue is of fundamental importance, since gravitational radiation is defined unambiguously if null infinity has a certain level of smoothness. However, there still exist several outstanding questions even on the analytical side, and we hope that ultimately our numerical work may shed further light on these.

The purpose of the present paper is to study the behaviour of the general conformal field equations near the cylinder representing $i^0$ and, in particular, near the critical sets $I^\pm$ where $\scri^\pm$ touches the cylinder $I$. We limit our discussion to two particular conformal representations: one in which $\scri$ `is diagonal', i.e., it is always transverse to the $t=\mathrm{const.}$ space-like hyper-surfaces; and the `horizontal representation', in which the $t=\mathrm{const}.$ hyper-surfaces tilt and become null for $t=\pm1$, coinciding with $\scri^\pm$. In an $(r,t)$ diagram this is reflected by horizontal $\scri^\pm$, as shown in Fig.~\ref{fig:cylinder}.

Our numerical system was chosen as simple as possible. Apart from the linearisation we also made use of the spherical symmetry of Minkowski space to decompose the spin-2 field representing the perturbed (rescaled) Weyl curvature into harmonic modes.

Using this system we have been able to demonstrate that in the diagonal representation we can reach the singular set $I^+$ at $t=1$ without loss of convergence in several test cases. This is not so surprising since the equations become singular `only' at $I^+$, i.e., $(r,t)=(0,1)$. With the explicit RK4 method that we use we never actually evaluate the equation at $t=1$ so we never see the singular coefficient $1-\kappa'(0)$ in front of $\del_t\phi_4$. Hence, the loss of hyperbolicity at $I^+$ does not seem to be important. However, as mentioned already in sect.~\ref{sec:code-test-exact} the break-down of hyperbolicity beyond $t=1$ manifests itself very quickly. 

We have explored several possibilities to go beyond the critical point $(t,r)=(1,0)$. These include a change of coordinates to put $\scri$ on a fixed grid location and the dropping of grid points which lie beyond $\scri$ outside the physical manifold. Both of these methods have their drawbacks and we will report on these attempts in a later paper.

The situation is different in the horizontal representation. Here we cannot reach $t=1$ because the equation for $\phi_4$ becomes singular for all values of $r$ and the characteristic speed of $\phi_4$ blows up at $t=1$. This has the consequence that running the code with any fixed time-step inadvertently leads to instabilities near $t=1$. We can get around this by using adaptive time-stepping with the consequence that we can get arbitrarily close to $t=1$ but in arbitrarily long wall time.

There are several questions which need to and will be addressed in the future:
In the diagonal representation we cannot compute far beyond the singular set. As we pointed out in sect.~\ref{sec:code-test-exact} the reason is due to the loss of hyperbolicity in every neighbourhood of $I^+$. Consequently, the code is very unforgiving and blows up very quickly. If we were able to successfully compute the numerical solutions beyond the critical sets, this would mean that we could choose a hyperboloidal surface and obtain the induced data from the numerical solution. It would then be possible to use those as initial data for other numerical hyperboloidal codes, for example \cite{huebner01:_from_now,huebner96:_numer_glob,Frauendiener:1998vo,Frauendiener:1998th,Frauendiener:2002ux,Moncrief:2009ds,Rinne:2010ew}. This would allow us to compute a large physically relevant portion of null infinity -- and therefore the gravitational wave forms -- in an unambiguous way.

In future work we intend to explore the various possibilities of numerically constructing hyperboloidal hyper-surfaces. Apart from the attempts to compute beyond the critical sets into the conformal manifold we have started to implement the equations without the mode decomposition and will report on this work elsewhere. 

Another aspect that needs to be explored is the use of a compact conformal representation such as the one which exhibits Minkowski space as a subset of the Einstein cylinder. This would allow us to eliminate the artificial boundary at $r=1$.

\section{Acknowledgments}
\label{sec:acknowledgments}

This work has been supported by Marsden grant UOO0922 of the Royal Society of New Zealand. The authors are grateful to H. Friedrich, B. Schmidt, O. Reula and M. Tiglio for extensive discussions of various aspects of this research.

\appendix

\section{An exact solution of the spin-2 equation}
\label{sec:an-exact-solution}

We very briefly describe how to derive an exact solution of the system~\eqref{eq:11}--\eqref{eq:12}. A detailed derivation can be found in~\cite{doulis2012:phdthesis}. The idea is to produce a solution on Minkowski space and then trace through the transformations described in sect.~\ref{sec:minkowski-space-time} to obtain a solution in the conformal representation where spatial infinity is represented as a cylinder.

The spin-2 system on Minkowski space with the metric $\tilde g$~\eqref{eq:1} using modal decomposition consists of the evolution system
\begin{equation}
  \label{eq:A1}
  \begin{aligned}
   R \, \del_T \tilde{\phi}_0 - R \, \del_R \tilde{\phi}_0 - \tilde{\phi}_0 + \alpha_2 \tilde{\phi}_1 &= 0 ,\\
   2R \, \del_T \tilde{\phi}_1 + 2 \tilde{\phi}_1 - \alpha_2 \tilde{\phi}_0 + \alpha_0 \tilde{\phi}_2 &= 0,\\
   2R \, \del_T \tilde{\phi}_2 - \alpha_0 (\tilde{\phi}_1 - \tilde{\phi}_3)  &= 0,\\
   2R \, \del_T \tilde{\phi}_3 - 2 \tilde{\phi}_3 - \alpha_0 \tilde{\phi}_2 + \alpha_2 \tilde{\phi}_4 &= 0,\\
   R \, \del_T \tilde{\phi}_4 + R \, \del_R \tilde{\phi}_4 + \tilde{\phi}_4 - \alpha_2 \tilde{\phi}_3 &= 0,
  \end{aligned}
\end{equation}
and the constraints
\begin{equation}
  \label{eq:A2}
  \begin{aligned}
   2R \, \del_R \tilde{\phi}_1 + 6 \tilde{\phi}_1 - \alpha_0 \tilde{\phi}_2 - \alpha_2 \tilde{\phi}_0 = 0,\\
   2R \, \del_R \tilde{\phi}_2 + 6 \tilde{\phi}_2 - \alpha_0 \tilde{\phi}_3 - \alpha_0 \tilde{\phi}_1 = 0,\\
   2R \, \del_R \tilde{\phi}_3 + 6 \tilde{\phi}_3 - \alpha_0 \tilde{\phi}_2 - \alpha_2 \tilde{\phi}_4 = 0.
  \end{aligned}
\end{equation}
The first two evolution equations and the first constraint equation can be used to obtain a 2nd order wave equation for $\phi_0$ alone:
\begin{equation}
  \label{eq:A3}
  R^2\,\del^2_T \tilde{\phi}_0 - R^2\,\del^2_R \tilde{\phi}_0 + 4\,R\,\del_T \tilde{\phi}_0 - 6\,R\,\del_R \tilde{\phi}_0 + 
  (\alpha^2_2 - 4) \tilde{\phi}_0= 0.
\end{equation}
With the ansatz 
\begin{equation*}
 \tilde{\phi}_0(T,R) \equiv \mathcal{R}(R)\, \mathcal{T}(T)
\end{equation*}
eq.~\eqref{eq:A3} takes the `almost' separable form 
\begin{equation*}
R\,\frac{\ddot{\mathcal{T}}}{\mathcal{T}} + 4\,\frac{\dot{\mathcal{T}}}{\mathcal{T}} = 
R\,\frac{\mathcal{R}''}{\mathcal{R}} + 6\,\frac{\mathcal{R}'}{\mathcal{R}} - \frac{\alpha^2_2 - 4}{R}, 
\end{equation*}
where $\dot{f}$ and ${f'}$ denote differentiation of a function $f$ with respect to the time and spatial 
coordinate, respectively. On the right hand side we have a function of $R$ only, while the left hand side is a linear polynomial in $R$ with $T$-dependent coefficients. Taking another $R$-derivative of this equation yields an equation between a function of $R$ on the right and a function of $T$ on the left. Thus, they must be constant. This leads to the ansatz:
\begin{equation} 
 \label{eq:A4}
 R\,\frac{\mathcal{R}''}{\mathcal{R}} + 6\,\frac{\mathcal{R}'}{\mathcal{R}} - \frac{\alpha^2_2 - 4}{R} = k\,R + 4 \omega,
\end{equation}
where the quantities appearing on the r.h.s must be arbitrary complex constants. Alternatively, the above ansatz can be interpreted as 
\begin{equation}
  \label{eq:A5}
  R\,\frac{\ddot{\mathcal{T}}}{\mathcal{T}} + 4\,\frac{\dot{\mathcal{T}}}{\mathcal{T}} = k\,R + 4 \omega.
\end{equation}
Obviously, one can set
\begin{equation*}
\frac{\ddot{\mathcal{T}}}{\mathcal{T}} = k, \quad   \frac{\dot{\mathcal{T}}}{\mathcal{T}} = \omega,
\end{equation*}
which entails that $k \equiv {\omega^2}$. Thus, the most general solution of \eqref{eq:A5} is of the form
\begin{equation}
  \label{eq:A6}
  \mathcal{T}(T) = e^{\omega\,T},
\end{equation}
up to an irrelevant scale factor. The equation for the spatial part also admits an analytic solution. This provides a solution for arbitrary $l$. In order to keep things as simply as possible, we will 
consider here only the first non-trivial case $l = 2$. For this choice of $l$ \eqref{eq:A4} 
admits a solution of the form\footnote{This solution corresponds to a special case of Teukolsky's family of quadrupole solutions~\cite{Teukolsky:1982jr} with $F(x)=\exp(\omega x)$. We thank an unknown referee for pointing us towards this paper.}
\begin{equation}
  \label{eq:A7}
 \mathcal{R}(R) =  \frac{1}{R^5} \left(e^{\omega R}(3 - 6\omega R  + 6\omega^2R^2 - 4 \omega^3R^3 + 2 \omega^4 R^4) \, c_1 - 3  e^{-\omega R}\,c_2\right).
\end{equation}
Using \eqref{eq:A6}, \eqref{eq:A7} and the first four evolution equations, it is straightforward to 
show that
\begin{equation}
  \label{eq:A8}
  \begin{aligned}
 \tilde{\phi}_0 &= \frac{1}{R^5} \left(e^{\omega(T+R)}(3 - 6 \omega R  + 6 \omega^2R^2 - 4 \omega^3R^3 + 2 \omega^4 R^4) \, c_1 - 3  e^{\omega(T-R)}\,c_2\right)\\
 \tilde{\phi}_1 &= \frac{1}{R^5} \left(e^{\omega(T+R)}(-6 + 9\omega R - 6 \omega^2R^2 + 2 \omega^3 R^3)\,c_1 +  e^{\omega(T-R)}(2 + \omega R)\,c_2\right)\\
 \tilde{\phi}_2 &= \frac{\sqrt6}{R^5} \left(e^{\omega(T+R)}(3 - 3\omega R + \omega^2 R^2)\,c_1 -  e^{\omega(T-R)}(3 + 3\omega R + \omega^2 R^2)\,c_2\right)\\
 \tilde{\phi}_3 &= \frac{1}{R^5} \left(e^{\omega(T+R)}(-2 + \omega R)\,c_1 +  e^{\omega(T-R)}(6 + 9\omega R + 6 \omega^2R^2 + 2 \omega^3 R^3)\,c_2\right)\\
 \tilde{\phi}_4 &= \frac{1}{R^5} \left(e^{\omega(T+R)} \, c_1 +  e^{\omega(T-R)}(3 + 6\omega R  + 6 \omega^2R^2 + 4 \omega^3R^3 + 2 \omega^4 R^4)\,c_2\right).
  \end{aligned}
\end{equation}  
Relations \eqref{eq:A8} provide a family of solutions for the system \eqref{eq:A1}--\eqref{eq:A2}, 
which satisfy a separation of variables ansatz. This family describes a superposition of ingoing and outgoing waves. Before we continue to the derivation of the 
solutions for the complete system \eqref{eq:11}--\eqref{eq:12}, we will make a specific 
choice for the constants $c_1, c_2, \omega$, namely $c_1 = \frac1{3}, c_2 = 0, \omega=0$. With 
this choice \eqref{eq:A8} reduces to a static solution:
\begin{equation}
  \label{eq:A9}
   \tilde{\phi}_0 = \frac{1}{R^5}, \quad
   \tilde{\phi}_1 = -\frac{2}{R^5}, \quad
   \tilde{\phi}_2 = \frac{\sqrt{6}}{R^5}, \quad
   \tilde{\phi}_3 = -\frac{2}{R^5}, \quad
   \tilde{\phi}_4 = \frac{1}{R^5}.
\end{equation}  
Using this static solution we now perform successively the transformations described above to obtain the corresponding solution with respect to the metric $g$, see eq.~\eqref{eq:5}. The main ingredient we have to be aware of is that according to~\cite{Penrose:1984wr} the spin-2 zero-rest-mass field transforms, under rescalings of the form \eqref{eq:4}, as $\phi_{ABCD}= \Theta^{-1} \, \tilde{\phi}_{ABCD}$. Using this formula and the appropriate rescalings of the spinor components and choosing the conformal representation with $\kappa(r) = r/(1+r)$ we, finally, end up with an exact solution of the complete system~\eqref{eq:11}--\eqref{eq:12}:
\begin{equation}
 \label{eq:A13}
  \begin{aligned}
    \phi_0(t,r) &= \frac{r^2 (1+r - t)^4}{(1+r)^7}, \\
    \phi_1(t,r) &= \frac{2 r^2 (1+ r - t)^3 (1+ r + t)}{(1 + r)^7}, \\
    \phi_2(t,r) &= \frac{\sqrt{6} r^2 (1+ r - t)^2 (1+ r + t)^2}{(1+ r)^7}, \\
    \phi_3(t,r) &= \frac{2 r^2 (1+ r - t) (1+ r + t)^3}{(1+ r)^7}, \\
    \phi_4(t,r) &= \frac{r^2 (1+ r + t)^4}{(1+ r)^7}.    
  \end{aligned} 
\end{equation}

\section{Spin-weighted spherical harmonics}
\label{sec:spin-spher-harm}
We briefly summarize the derivation and conventions of the spin-weighted spherical harmonics $\Y{s}{lm}$ from \cite{Penrose:1984wr} and compare them to Friedrich's definition of the functions ${{T_i}^j}_k$, see for example \cite{Friedrich:1998tc}. 

Let $V^l$ be the $(l+1)$-dimensional vector space of complex homogeneous polynomials of degree $l$ in two complex variables $z_1$ and $z_2$, i.e.\ the vector space spanned by the basis $\{\phi^l_0,\ldots,\phi^l_l\}$ where $\phi^l_k(z_1,z_2):=z_1^k z_2^{l-k}$. Let a sesquilinear form be given on $V^l$ by
\[\left<\phi^l_k,\phi^l_m\right>:=k! (l-k)! \delta_{km},\]
where our convention is that this form is anti-linear in the second argument; this is a scalar product on $V^l$.
For every $t\in SU(2)$  in the standard matrix representation  
\begin{equation}
\label{eq:SU2}
t=
\begin{pmatrix}
  {t^{0}}_{0} & {t^{0}}_{1}\\
{t^{1}}_{0} & {t^{1}}_{1}
\end{pmatrix},
\end{equation}
with complex numbers  ${t^{0}}_{0}$, ${t^{0}}_{1}$, ${t^{1}}_{0}$ and ${t^{1}}_{1}$ satisfying
\[{t^{0}}_{0} {t^{1}}_{1}- {t^{0}}_{1} {t^{1}}_{0}=1,\quad  {t^{0}}_{0}= \overline {{t^{1}}_{1}},\quad
  {t^{0}}_{1}=- \overline{{t^{1}}_{0}},\]
we consider the map 
\[T_t: \mathbb C^2\rightarrow \mathbb C^2,\quad (z_0,z_1)^T\mapsto t\cdot (z_0,z_1)^T,
\] 
where ``$\cdot$'' denotes a matrix product. For every non-negative integer $l$, this induces a representation $U^l$ of $SU(2)$ on $V^l$ as follows
\[U^l: SU(2)\times V^l\rightarrow V^l,\quad (t,\phi)\mapsto U^l_t\phi:=\phi\circ T_{t^{-1}}.\]
It is shown in \cite{Sugiura:1990vj} that each of these representations is unitary (up to normalizations with respect to the above scalar product, see below) and irreducible. Indeed every irreducible unitary representation of $SU(2)$ is equivalent to $U^l$ for some choice of $l$. 

For every $l=0,1,\ldots$, and $m,k=0,\ldots,l$, and $t\in SU(2)$, we find 
\[\left<\phi^l_m,U^l_t\phi^l_k\right>
=m! (l-m)!
\begin{pmatrix}
  l\\m
\end{pmatrix}
{{t^{(a_1}}_{(b_1}{t^{a_2}}_{b_2}\cdots {t^{a_l)_m}}_{b_l)_k}},
\]
where the notation $(a_1 a_2\cdots a_l)_m$ means the following: (i) symmetrize over the indices $a_1,\ldots,a_l$, (ii) replace $m$ indices by $0$ and $l-m$ indices by $1$. A normalization yields the matrix elements of the unitary representations of $SU(2)$; those are Friedrich's functions
\begin{align}
{{T_l}^m}_k(t)&:=\frac{1}{\sqrt{m!(l-m)!}}\frac{1}{\sqrt{k!(l-k)!}}\left<\phi^l_m,U^l_t\phi^l_k\right>\notag\\
\label{eq:defTijk}
&=\sqrt{
  \begin{pmatrix}
    l\\m
  \end{pmatrix}
  \begin{pmatrix}
    l\\k
  \end{pmatrix}}\,\,{t^{(a_1}}_{(b_1}{t^{a_2}}_{b_2}\cdots {t^{a_l)_m}}_{b_l)_k}.
\end{align}
According to the Peter-Weyl theorem \cite{Sugiura:1990vj} the functions $\sqrt{l+1}\, {{T_l}^m}_k$, with $l=0,1,2,\ldots$, and $m,k=0,1,\ldots,l$ form an orthonormal basis of the Hilbert space $L^2(SU(2))$ with respect to the standard normalized Haar measure. Hence, every function on $SU(2)$ can be expanded in this basis.

Now, for any non-negative integer\footnote{All of the following constructions also work for half-integer values of $l$. In this case, the other indices $m$ and $s$ used in the following must also be half-integer valued. In this paper, we focus exclusively on the integer case.} $l$, let the space $W^{2l}$ be the $(2l+1)$-dimensional vector space of those spinorial tensor fields on Minkowski space which are spanned by the basis elements
\[Z(l,m)_{A_1\ldots A_{l-m} B_1\ldots B_{l+m}}
:=\hat o_{(A_1}\cdots \hat o_{A_{l-m}}\hat\iota_{B_1}\cdots\hat\iota_{B_{l+m})},\]
for $m=-l,\ldots,l$, see Eq.~(4.15.93) in \cite{Penrose:1984wr}.  Here $(\hat o^A, \hat \iota^A)$ is the ``constant dyad'' of Penrose (as opposed to a ``rotated dyad'' below) which is normalized by $1=\hat o_A\hat \iota^A$. Clearly, there is an isomorphism between $\Psi: W^{2l}\rightarrow V^{2l}$ with
\[\Psi:Z(l,m)_{A_1\ldots A_{2l}}\mapsto Z(l,m)_{A_1\ldots A_{2l}}\Pi^{A_1}\ldots \Pi^{A_{2l}}=\phi^{2l}_{l-m},\]
for
\[\Pi^A:=-z_1\hat o^A+z_0\hat\iota^A.\]
Roughly speaking, each $\hat o_A$ is replaced by $z_0$ and each $\hat \iota_A$ by $z_1$. By using this isomorphism, we can induce a scalar product on $W^{2l}$ from the one above on $V^{2l}$ (in the following we do often not write spinor indices when it is convenient):
\[\left<Z(l,m),Z(l,k)\right>:=\left<\Psi(Z(l,m)),\Psi(Z(l,k))\right>=(l-k)! (l+k)! \delta_{km}.\]
However, we also have a \textit{geometric} scalar product on this space of space spinors in Minkowski space constructed as follows. Let
\[T^{A A'}:=\frac {1}{\sqrt 2}(\hat o^A \hat o^{A'}+\hat\iota^A\hat\iota^{A'}).\]
Then we set
\[\left<Z(l,m),Z(l,k)\right>_G
:=T^{A_1 A_1'}\cdots T^{A_{2l} A_{2l}'}Z(l,m)_{A_1\ldots A_{2l}}\overline{Z(l,k)}_{A_{1}'\ldots A_{2l}'}.\]
Noting that $T^{A A'}\hat o_A=\hat \iota^{A'}/\sqrt 2$ and $T^{A A'}\hat\iota_A=-\hat o^{A'}/\sqrt 2$, we find
\[\left<Z(l,m),Z(l,k)\right>_G=\frac{(l-k)!(l+k)!}{2^l(2l)!}\delta_{mk}
=\frac{1}{2^l(2l)!}\left<Z(l,m),Z(l,k)\right>;\]
cf.\ Eq.~(4.15.94) in \cite{Penrose:1984wr}. 

We can also use the isomorphism $\Psi$ and the representations $U^{2l}$ of $SU(2)$ on $V^{2l}$ above to define representations of $SU(2)$ on $W^{2l}$, which we also call $U^{2l}$, as follows:
\[U^{2l}: SU(2)\times W^{2l}\rightarrow W^{2l}, \quad
(t,Z(l,m))\mapsto U^{2l}_t Z(l,m):= \Psi^{-1}(U^{2l}_t\Psi(Z(l,m))).\]
We can write this as
\[ U^{2l}_t Z(l,m) _{A_1\ldots A_{l-m} B_1\ldots B_{l+m}}=o_{(A_1}\cdots  o_{A_{l-m}}\iota_{B_1}\cdots\iota_{B_{l+m})},\]
where
\begin{equation}
  \label{eq:SU2action}
  o_A=\overline{{t^0}_0}\hat o_A+\overline{{t^1}_0}\hat \iota_A, \quad \iota_A=\overline{{t^0}_1}\hat o_A+\overline{{t^1}_1}\hat \iota_A.
\end{equation}

In (4.15.95)  of \cite{Penrose:1984wr}, one defines the functions
\[{}_sZ_{l,m}(t):=Z(l,m)_{A_1\ldots A_{l-m} A_{l-m+1}\ldots A_{2l}}\iota^{(A_1}\cdots \iota^{A_{l-s}}o^{A_{l-s+1}}\cdots o^{A_{2l})},\]
on $SU(2)$ (the quantity $W$ in (4.15.95) in \cite{Penrose:1984wr} is identically unity in our case of pure rotations). Noting that $T^{A A'}o_A=\iota^{A'}/\sqrt 2$ and $T^{A A'}\iota_A=-o^{A'}/\sqrt 2$, we can write this as
\begin{align*}
  {}_sZ_{l,m}(t)&=(-1)^{m+2} 2^lT^{A_1 A_1'}\cdots T^{A_{2l} A_{2l}'} Z(l,m)_{A_1\ldots A_{2l}} 
  \overline{U^{2l}_tZ(l,s)}_{A'_1\ldots A'_{2l}}\\
  &=(-1)^{l+s} 2^l\left<Z(l,m), U^{2l}_tZ(l,s)\right>_G
=\frac{(-1)^{l+s}}{(2l)!}\left<Z(l,m), U^{2l}_tZ(l,s)\right>\\
&=\frac{(-1)^{l+s}}{(2l)!}\left<\phi^{2l}_{l-m}, U^{2l}_t \phi^{2l}_{l-s}\right>
\end{align*}
It follows
\[  {}_sZ_{l,m}(t)=\frac{(-1)^{l+s}}{(2l)!}\,{{T_{2l}}^{l-m}}_{l-s}(t)\,\sqrt{(l-m)!(l+m)!(l-s)!(l+s)!}.\]
The spin-weighted spherical harmonics $\Y{s}{lm}$ are defined by (4.15.98) in \cite{Penrose:1984wr} as normalizations of the functions ${}_sZ_{l,m}(t)$:
\[\Y{s}{lm}(t):=(-1)^{l+m}{}_sZ_{l,m}(t)\sqrt{\frac{(2l+1)!(2l)!}{4\pi (l-m)!(l+m)!(l-s)!(l+s)!}}.\]
Hence, it follows 
\begin{equation}
  \label{eq:relationYlmTijk}
  \Y{s}{lm}(t)=(-1)^{s+m}\sqrt{\frac{2l+1}{4\pi}}\,\,\,{{T_{2l}}^{l-m}}_{l-s}(t).
\end{equation}
We see explicitly that these functions are defined for every $l=0,1,\ldots$ and for every $m$ and $s$ constrained to $m=-l,\ldots,l$ and $s=-l,\ldots,l$.

The reader should notice that, while the representation spaces $V^l$ for $l=0,1,\ldots$ give rise to \textit{all} unitary irreducible representation of $SU(2)$ and hence to a basis of the Hilbert space $L^2(SU(2))$ as above, the representation spaces $W^{2l}$ (which are isomorphic to $V^{2l}$) for integer-valued $l$ give rise to only ``half'' of the irreducible representations. In particular, it follows that the functions $\Y{s}{lm}$ with integer-valued indices $l$, $m$ and $s$ do \textit{not} form a basis of $L^2(SU(2))$. However,  for a fixed value of $s$, the functions $\Y{s}{lm}$ form a basis of functions with spin-weight $s$  in $L^2(SU(2))$. This is not only true for the integer spin case, which we have considered here, but also for the half-integer spin case.

So far, we have considered $\Y{s}{lm}$ as functions on $SU(2)$, and not, as it is usually done, as functions on $\mathbb S^2$. We argue as follows without going into the details (which can be found for instance in \cite{Beyer:2011uz}). It is a fact that $SU(2)$ is diffeomorphic to $\mathbb S^3$, and $\mathbb S^3$ is the Hopf bundle over $\mathbb S^2$ with the Hopf map $\pi$ as the bundle projection. This bundle is not trivial and hence smooth functions on $SU(2)$ do in general not give rise to smooth functions on $\mathbb S^2$. We can, however,
choose a smooth local section in the bundle, i.e.\ a smooth map $\sigma:U\rightarrow SU(2)$ compatible with $\pi$ where $U$ is an open subset of $\mathbb S^2$. With this the smooth functions $\Y{s}{lm}$ can be pulled back to smooth functions $\Y{s}{lm}\circ\sigma$ defined on $U$. The choice of local section made by Penrose is the following. We introduce standard polar coordinates $(\theta,\phi)$ on $\mathbb S^2$ and set $U:=\mathbb S^2\backslash\{\theta=0,\pi\}$. Then, the following defines a smooth local section:
\[\sigma:U\rightarrow SU(2),\quad (\theta,\phi)\mapsto t=
\begin{pmatrix}
  {t^{0}}_{0} & {t^{0}}_{1}\\
{t^{1}}_{0} & {t^{1}}_{1}
\end{pmatrix}
=
\begin{pmatrix}
  e^{-i \phi/2}\cos\frac\theta2 &  -e^{-i \phi/2}\sin\frac\theta2\\
  e^{i \phi/2}\sin\frac\theta2 &  e^{i \phi/2}\cos\frac\theta2
\end{pmatrix}.
\]
The reader should compare this to Eq.~(4.15.123) in \cite{Penrose:1984wr}:
\[
\begin{pmatrix}
 - {t^{0}}_{1} &  {t^{0}}_{0}\\
   -{t^{1}}_{1} &  {t^{1}}_{0}
\end{pmatrix}
=\begin{pmatrix}
  \hat o_A o^A & \hat o_A \iota^A\\
  \hat \iota_A o^A & \hat \iota_A \iota^A
\end{pmatrix}
=
\begin{pmatrix}
  e^{-i \phi/2}\sin\frac\theta2 &  e^{-i \phi/2}\cos\frac\theta2\\
  -e^{i \phi/2}\cos\frac\theta2 &  e^{i \phi/2}\sin\frac\theta2
\end{pmatrix},
\]
which is consistent.
Using these coordinate expressions, we can compute the functions ${{{T_l}^m}_k}$ as functions on $\mathbb S^2$ via Eq.~\eqref{eq:defTijk} and then $\Y{s}{lm}$ by Eq.~\eqref{eq:relationYlmTijk}. It is important to notice that while ${{{T_l}^m}_k}$ and $\Y{s}{lm}$ are smooth functions on $SU(2)$ and their pull-backs to $U$ are smooth, there is in general no smooth (not even a continuous) extension from the dense subset $U$ to $\mathbb S^2$.
Here are some examples of explicit expressions for some harmonics
\begin{equation*}
  \Y{0}{1,-1}=\sqrt{\frac{3}{8\pi}}\sin\theta e^{-i\phi},\quad \Y{0}{1,0}=\sqrt{\frac{3}{4\pi}}\cos\theta,
  \quad \Y{0}{1,1}=-\sqrt{\frac{3}{8\pi}}\sin\theta e^{i\phi},
\end{equation*}
and
\begin{gather*}
  \Y{2}{2,-2}=\sqrt{\frac{5}{4\pi}}\sin^4\frac{\theta}{2} e^{-2i\phi},\quad
\Y{2}{2,-1}=\sqrt{\frac{5}{4\pi}}\sin^2\frac{\theta}{2}\sin\theta e^{-i\phi},\\
\Y{2}{2,0}=\sqrt{\frac{15}{32\pi}}\sin^2\theta,\\
\end{gather*}
etc.

\providecommand{\newblock}{}

\end{document}